\documentclass[11pt]{article}

\usepackage[top=60pt,bottom=70pt,left=75pt,right=75pt]{geometry}

\usepackage{amsmath, amssymb, amsfonts, latexsym, euscript, mathalfa, mathrsfs}
\usepackage{graphicx, color, epsfig, xcolor }

\usepackage{setspace, sectsty}

\usepackage{braket, cancel}
\usepackage{tensor}

\usepackage{jheppub} 

\newcommand{\nn}{\nonumber}

\def\beq{\begin{equation}}
\def\eeq{\end{equation}}
\newcommand{\be}{\begin{equation}}
\newcommand{\ee}{\end{equation}}
\newcommand{\dd}{d}
\newcommand{\me}{e}
\newcommand{\Real}{\text{Re}}
\newcommand{\Imag}{\text{Im}}
\newcommand{\ii}{i}
\newcommand{\prim}{\nu_2}
\newcommand{\ls}{\ell_s}
\newcommand{\vol}{\text{vol}}
\newcommand{\e}{\mathrm{e}}

\begin{document}

% format
\baselineskip=18pt  % a la harvmac
\numberwithin{equation}{section}  % make eq labels (sec.num)
\allowdisplaybreaks  % allow page breaks in displayed eqs

\begin{titlepage}
\vspace*{-1cm}
\begin{center}
 {\Large\bf On Type IIA AdS$_3$ solutions and massive GK geometries}
\end{center}
\vspace*{-.5cm}
%
%% ========== title (note version) ends here ==========

%% ========== title (paper version, a la harvmac) begins here ==========
\begin{center}
{
Christopher Couzens$^{a}$ \footnote{cacouzens@khu.ac.kr},  Niall T. Macpherson$^{b,c}$ \footnote{ntmacpher@gmail.com}, 	Achilleas Passias$^{d}$ \footnote{achilleas.passias@gmail.com}\\[4mm]}
\vspace{5mm}
$a$: Department of Physics and Research Institute of Basic Science,\\
  Kyung Hee University, Seoul 02447, Republic of Korea\\
  
\vskip 3mm
$b$: Departamento de F\'isica de Part\'iculas\\
 Universidade de Santiago de Compostela\\
	and\\
	$c$: Instituto Galego de F\'isica de Altas Enerx\'ias (IGFAE)\\
	R\'ua de Xoaqu\'in D\'iaz de R\'abago s/n\\
	E-15782 Santiago de Compostela, Spain

\vskip 3mm
$d$: Department of Nuclear and Particle Physics, 
Faculty of Physics, National and Kapodistrian University of Athens, 
Athens 15784, Greece \\

\vspace*{.5cm}
\end{center}

\abstract

We give the necessary and sufficient conditions for warped AdS$_3$ (and Mink$_3$) solutions of Type II supergravities to preserve ${\cal N}=(2,0)$ supersymmetry, in terms of geometric conditions on their internal space M$_7$. Such solutions possess a canonical ten-dimensional Killing vector that can be either time-like or null. In this work we classify the null case in massive Type IIA supergravity which necessitates that M$_7$ decomposes as a circle fibration over a six-dimensional base with orthogonal SU(2)-structure containing a complex four-manifold.  We narrow our focus to solutions for which M$_7$ becomes $\mathbb{T}^2$ fibred over a foliation of a K\"{a}hler manifold over an interval. We find a class of solutions which are the massive Type IIA version of GK geometries and present an extremal problem which computes the central charge of the solution using just topology. Finally, we present geometric conditions for AdS$_3$ solutions to preserve arbitrary extended chiral supersymmetry.

\noindent

\end{titlepage}
%\newpage
%%%%%%%%%%%%%%%%%%%%%%%%%%%%%%%%%%%%%%%%%%%
%%%           TITLE ENDS HERE
%%%%%%%%%%%%%%%%%%%%%%%%%%%%%%%%%%%%%%%%%%%

\tableofcontents
%\printindex

%%%%%%%%%%%%%%%%%%%%%%%%%%%%%%%%%%%%%%%%%%%
%%%        MAIN TEXT BEGINS HERE
%%%%%%%%%%%%%%%%%%%%%%%%%%%%%%%%%%%%%%%%%%%

\section{Introduction}

Two-dimensional conformal field theories (CFTs) hold a special place in the landscape of CFTs. They feature prominently in string theory, describing the world-sheet dynamics of strings and have been studied extensively in the literature. As they admit an infinite dimensional conformal algebra they are heavily constrained and in certain cases completely solvable. When 2d CFTs preserve (at least) $\mathcal{N}=(2,0)$ supersymmetry, $c$-extremization \cite{Benini:2012cz,Benini:2013cda} computes the central charge and R-charges (thus also conformal dimensions of certain operators) of the strongly coupled IR fixed point using only UV data. The key observation is that 2d $\mathcal{N}=(2,0)$ SCFTs have a U$(1)$ R-symmetry. Though the R-symmetry may mix with flavour symmetries along the RG flow, the exact R-symmetry in the IR extremizes the central charge viewed as a functional of possible R-symmetry choices, at the IR fixed point. Thus, from knowing just UV data, and with some mild assumptions in tow, one can obtain information about the IR fixed point. This is a direct 2d analogue of $a$-maximisation for 4d SCFTs \cite{Intriligator:2003jj}.

As one may expect given AdS/CFT, in gravity, there are geometric extremal problems dual to the field-theoretic ones. The geometric dual of $a$-maximization was derived in \cite{Martelli:2005tp,Martelli:2006yb} for AdS$_5\times$SE$_5$ geometries and the geometric dual of $c$-extremization was found in \cite{Couzens:2018wnk} for so called ``GK" geometries \cite{Kim:2005ez,Kim:2006qu,Gauntlett:2007ts}. The geometric extremal problem was later extended in \cite{vanBeest:2020vlv} to the F-theoretic extension of GK geometries of \cite{Couzens:2017nnr}.  Further advances in the geometric dual of $c$-extremization for the GK geometry class have been made in \cite{Kim:2019umc,Gauntlett:2019pqg,Hosseini:2019ddy,Hosseini:2019use,Gauntlett:2019roi,Gauntlett:2018dpc}. 

However, whereas $a$-maximization and $c$-extremization work for generic field theories obeying certain mild assumptions, the geometric extremal problems are only defined for certain classes of solutions and there are holographic SCFTs whose duals are not contained within these classes. 
It is natural to conjecture that there is an extremal problem for any AdS$_3$ solution in supergravity with at least $\mathcal{N}=(2,0)$ supersymmetry which is the geometric dual of $c$-extremization for the putative dual field theory.\footnote{Similar comments of course apply for any AdS solution where there is a dual field-theoretic extremal problem. }
It is therefore an interesting problem to extend these geometric extremal problems to cover the full complement of holographic SCFTs. One of the key results needed for progress on these geometric extremal problems was a thorough understanding of the underlying geometry of the system. For \cite{Couzens:2018wnk} the underlying geometries are GK geometries which were first studied in \cite{Kim:2005ez} and arise from classifying AdS$_3$ solutions of Type IIB supergravity with 5-form flux and an SU$(3)$-structure. Therefore, extending the classification of all AdS solutions preserving fixed amounts of supersymmetry is a necessary requirement for making progress in extending these geometric extremal problems to further classes of geometries.\footnote{A complementary approach to these geometric duals studies the extremal problem from gauged supergravity, see \cite{Tachikawa:2005tq} for $a$-maximization and \cite{Karndumri:2013iqa} for $c$-extremization. From this perspective it is also interesting to classify supersymmetric AdS$_3$ solutions and then to obtain consistent truncations which uplift on these geometries. }
There has been a lot of interest in classifying AdS$_3$ solutions with various amounts of supersymmetry \cite{Kim:2005ez,Martelli:2003ki,Tsimpis:2005kj,Kim:2007hv,Figueras:2007cn,Donos:2008hd,OColgain:2010wlk,DHoker:2008lup,Estes:2012vm,Bachas:2013vza,Jeong:2014iva,Lozano:2015bra,Kelekci:2016uqv,Couzens:2017way,Eberhardt:2017uup,Dibitetto:2018iar,Dibitetto:2018ftj,Macpherson:2018mif,Legramandi:2019xqd,Lozano:2019emq,Lozano:2019jza,Lozano:2019zvg,Lozano:2019ywa,Couzens:2019mkh,Couzens:2019iog,Passias:2019rga,Lozano:2020bxo,Farakos:2020phe,Couzens:2020aat,Faedo:2020nol,Dibitetto:2020bsh,Passias:2020ubv,Faedo:2020lyw,Legramandi:2020txf,Zacarias:2021pfz,Emelin:2021gzx,Couzens:2019wls,Couzens:2021veb,Couzens:2021tnv,Macpherson:2021lbr,Macpherson:2022sbs}\footnote{See \cite{Filippas:2019ihy,Filippas:2020qku,Speziali:2019uzn,Rigatos:2020igd,Eloy:2020uix} for examples of further studies.} yet the analysis is still incomplete and there are interesting AdS$_3$ geometries still to be classified and constructed; this paper tightens the noose on classifying all AdS$_3$ solutions in Type II supergravity.

Another reason for interest in AdS$_3$ solutions stems from their presence in the near-horizon limit of black strings. The power of the near-horizon geometry is that many of the interesting observables of the black string may be obtained from the near-horizon rather than the asymptotic geometry. The entropy, angular momentum, electric and magnetic charges of a black string can all be obtained this way. However, the electrostatic potential and angular velocity require UV knowledge which is washed out in the near-horizon limit. As such if one restricts to understanding any of the first class of observables one needs only the near-horizon limit of the black string and not the full interpolating flow. In addition, the 2d CFT dual to the AdS$_3$ near-horizon gives a microscopic description for the Bekenstein--Hawking entropy of the black strings.\footnote{There can be some subtleties regarding the contribution of ``hair'' to the entropy, which the 2d SCFT does not see, however this is a subleading contribution to the entropy and can therefore be neglected in most examples. } The first example of computing the microstates of a black string using the near-horizon geometry was performed in the mid 90's by Strominger and Vafa \cite{Strominger:1996sh}. This was later extended to the MSW string \cite{Maldacena:1997de} in M-theory and in F-theory in \cite{Vafa:1997gr}. More recently there have been many advancements in studying the near-horizon of black strings. 
There has been recent interest in black strings with non-constant curvature horizons: in \cite{Boido:2021szx,Ferrero:2020laf,Hosseini:2021fge} black strings with spindle horizons were investigated whilst in \cite{Couzens:2021tnv,Suh:2021ifj} black strings with disc horizons were studied. In a tangential direction, there have been further advancements in studying black strings in F-theory probing various four-dimensional asymptotically flat spaces \cite{Bena:2006qm,Haghighat:2015ega,Grimm:2018weo,Couzens:2020aat,Couzens:2019wls} using their AdS$_3$ near-horizon geometries. 

AdS$_3$ vacua are also interesting when considering whether AdS vacua can be parametrically scale separated. One can construct minimally supersymmetric AdS$_3$ vacua in Type II supergravity of the form AdS$_3\times$M$_7$ with M$_7$ a $G_2$ manifold, see for example \cite{Farakos:2020phe,Emelin:2021gzx,Apers:2022zjx}. Since the CFT dual is two-dimensional and benefits from the infinite dimensional conformal symmetry, one may hope that this extra control allows one to answer this question conclusively. With suitable projection conditions imposed on the $\mathcal{N}=(2,0)$ solutions discussed in this work one is able to obtain solutions preserving only $\mathcal{N}=(1,0)$ supersymmetry which may be candidates for solutions with scale separation. It would be interesting to address this point in the future. 

The layout of the paper is as follows:
~\\
In section \ref{sec:SUSY} we present necessary and sufficient conditions for a warped AdS$_3\times $M$_7$ solution of Type II supergravity to preserve ${\cal N}=(2,0)$ supersymmetry in terms of geometric conditions on the internal space M$_7$. These conditions depend on the inverse AdS radius $m$: For $m\neq 0$ we find that M$_7$ necessarily decomposes as a U(1) fibration over a six-dimensional base M$_6$, with the U(1) realising the R-symmetry of the superconformal group OSp(2$|$2) as expected. When $m=0$ conditions for warped ${\cal N}=2$ three-dimensional Minkowski (Mink$_3$) vacua are recovered; while these still generically contain a U(1) isometry, this is no longer an R-symmetry and for restricted classes need to be present. For AdS$_3$ solutions specifically ($m\neq 0$) we find that solutions fall into two classes depending on whether a canonical ten-dimensional Killing vector is time-like or null. This section is supplemented by the technical appendices \ref{sec:AdS3app} and \ref{app:(2,0)}.

In section \ref{sec:null classification}, and for the rest of the paper, we narrow are focus to solution in massive Type IIA supergravity supporting a null Killing vector. For such solutions  M$_6$ necessarily supports an orthogonal SU(2)-structure, decomposing in terms of a complex vector and a four-manifold which in this case is complex. We introduce local coordinates for the complex vector and perform an analysis in terms of the SU(2)-structure torsion classes, reviewed in appendix \ref{app:torsions}.

In section \ref{sec:Additional symmetries}, in order to find solutions, we impose the existence of an additional Killing vector. This allows us to further refine the necessary and sufficient conditions for a supersymmetric solution. Inserting an ansatz for the SU$(2)$-structure manifold of the form of the product of two warped Riemann surfaces we find three classes of solutions. Buoyed by finding explicit solutions, in section \ref{sec:GenKahler} we study a more general ansatz for the SU$(2)$-structure manifold consisting of a warped K\"ahler manifold. We find a class of geometries which are the massive extension of GK geometries in five dimensions. These solutions are determined by the same master equation as GK geometries but contain D8-branes. In section \ref{sec:extremal} we show that one can define an extremal problem for determining the central charge of the solution using just the topology. This is the first example of an extremal problem for solutions of massive Type IIA supergravity. 

Finally, in section \ref{sec:extendedsusy} we use the  ${\cal N}=(2,0)$ supersymmetry conditions in appendix \ref{app:(2,0)}  to derive necessary and sufficient conditions  for AdS$_3$ solutions of Type II supergravity to preserve arbitrary extended chiral supersymmetry. We find that an ${\cal N}=(n,0)$ solution for $n\geq 2$ necessarily comes equipped with an anti-symmetric matrix of Killing vectors that should span the R-symmetry of whatever superconformal algebra a solution realises (there are many options \cite{Beck:2017wpm}). The  various $d=2$ superconformal algebras can be classified in terms of their R-symmetry and an associated representation \cite{Fradkin:1992bz}.\footnote{Strictly speaking this applies to the simple Lie superalgebras, but as these are in one-to-one correspondence with the algebras embeddable into supergravities with an AdS$_3$ factor, this subtelty is unimportant.} We make a conjecture that precisely relates the R-symmetry and representation to the anti-symmetric matrix of Killing vectors.

\section{Supersymmetry equations}\label{sec:SUSY}
We consider a bosonic background of Type II supergravity that 
preserves the symmetries of three-dimensional anti-de Sitter spacetime AdS$_3$.
The ten-dimensional metric takes the form of a warped product of a metric on
AdS$_3$ and a Riemanian metric on a seven-dimensional manifold M$_7$:
\beq
ds^2= e^{2A}ds^2(\text{AdS}_3)+ds^2(\text{M}_7)\, ,
\eeq
where the warp factor $e^{2A}$ is a function on M$_7$ and we assume that $\text{Ricci}(\text{AdS}_3)=-2m^2 g(\text{AdS}_3)$. The NS--NS three-form, $H^{(10{\rm d})}$ and the R--R fluxes $F$ take the form
\beq
H^{(10{\rm d})}= e^{3A}h_0 \text{vol}(\text{AdS}_3)+ H\, ,~~~ F= f_{\pm}+ e^{3A}\text{vol}(\text{AdS}_3)\wedge\star_7\lambda(f_{\pm})\, .
\eeq
Here, $H$ has support on M$_7$, and the Bianchi identity for $H^{(10{\rm d})}$ enforces $e^{3A} h_0$ to be a constant. $F$ is a polyform, the sum of the $p$-form R--R field-strengths with $p$ even for Type IIA supergravity and $p$ odd for Type IIB. The forms $f_\pm$ have support on M$_7$, with the upper sign corresponding to Type IIA and the lower to Type IIB. In particular, $f_+ = \sum_{p=0}^3 f_{2p}$ and $f_- = \sum_{p=0}^3 f_{2p+1}$. The operator $\lambda$ acts on the $p$-component of $f_\pm$
as
\begin{equation}
\lambda(f_\pm|_p) = (-1)^{\lfloor \frac{p}{2} \rfloor} f_\pm\bigg|_p\, ,
\end{equation}
and $\star_7$ is the Hodge operator involving the metric on M$_7$. Finally, the dilaton $\Phi$ is a function on M$_7$.

The conditions stemming from requiring that this background preserves $\mathcal{N}=(2,0)$ supersymmetry are derived in appendix \ref{app:(2,0)}. They involve two doublets of Spin$(7)$ Majorana spinors, $\chi_1^I$ and $\chi_2^I$, $I=1,2$ under the R-symmetry $\mathfrak{so}(2) \simeq \mathfrak{u}(1)$ of the supersymmetry algebra.
Without loss of generality (see appendix) they are taken to satisfy
\beq
\chi^{I\dag}_1\chi^{J}_1+\chi^{I\dag}_2\chi^{J}_2= 2e^{A}\delta^{IJ}\, ,~~~~\chi^{I\dag}_1\chi^{J}_1-\chi^{I\dag}_2\chi^{J}_2= c e^{-A}\delta^{IJ}\, ,
\eeq
for $c$ a constant and $\delta^{IJ}$ the Kronecker delta.

The supersymmetry constraints can then be expressed in terms of two 1-forms and a set of bi-spinors. The one 1-forms are given by
\begin{align}
\xi&= -i(\chi^1_1\gamma_a\chi^2_1 \mp  \chi^1_2\gamma_a\chi^2_2)\e^a\, ,~~~\tilde\xi= -i(\chi^1_1\gamma_a\chi^2_1 \pm  \chi^1_2\gamma_a\chi^2_2)\e^a\, ,~~~~~\langle \xi,\tilde{\xi}\rangle= 2c\, ,
\end{align}
where $\gamma^a$, $a = 1,2,\dots, 7$ are the generators of Cliff(7) and $\e^a$ gives an orthonormal frame on $T^*$M$_7$. Note that the 1-form $\xi$ cannot be set to zero globally when $m\neq 0$ without reducing supersymmetry to ${\cal N}=(1,0)$. The set of bi-spinors is
\begin{align}
\Psi^{IJ}&\equiv \chi_1^I\otimes \chi_2^{J\dag}=\frac{e^{A}}{2}\left(\delta^{IJ}\Psi^{(0)}+\sigma_1^{IJ}\Psi^{(1)}+i \sigma_2^{IJ}\Psi^{(2)}+\sigma^{IJ}_3\Psi^{(3)}\right),
\end{align}
where we have expanded in terms of $\delta^{IJ}$ and the Pauli matrices $(\sigma_1, \sigma_2,\sigma_3)$. They can be further decomposed into even, denoted by a plus subscript, and odd, denoted by a minus subscript, real parts as
\begin{equation}
\Psi^{(0,1,2,3)}=\Psi^{(0,1,2,3)}_++i \Psi^{(0,1,2,3)}_-\, .
\end{equation}
The supersymmetry constraints are
\begin{subequations}
\begin{align}
&d\tilde\xi= \iota_{\xi}H~~~~e^{3A}h_0=-m c\, ,\label{eq:bps0}\\[2mm]
&d_{H}(e^{2A-\Phi} (\Psi^{(1)}_{\mp}+i \Psi^{(3)}_{\mp}))=0\, ,\label{eq:bps1}\\[2mm]
&d_{H}(e^{3A-\Phi}(\Psi^{(1)}_{\pm}+i \Psi^{(3)}_{\pm}))\mp 2m e^{2A-\Phi}(\Psi^{(1)}_{\mp}+i \Psi^{(3)}_{\mp})=0\, ,\label{eq:bps2}\\[2mm]
&d_{H}(e^{3A-\Phi}\Psi^{(0)}_{\pm})\mp 2m e^{2A-\Phi}\Psi^{(0)}_{\mp}=\frac{1}{4}e^{3A}\star_7 \lambda f_{\pm}\, ,\label{eq:bps3}\\[2mm]
&d_{H}(e^{A-\Phi}\Psi^{(2)}_{\pm})=\frac{1}{8}(\tilde \xi\wedge +\iota_{\xi})f_{\pm}\, ,\label{eq:bps4}\\[2mm]
&e^{A}(\Psi^{(0)}_{\mp},f_{\pm})_7=\mp \left(m+\frac{1}{2} e^{-A}ch_0\right) e^{-\Phi}\text{vol}(\text{M}_7), .\label{eq:bps5}
\end{align}
\end{subequations}
Note there are further conditions implied by this, for instance acting on \eqref{eq:bps4} with $\tilde \xi\wedge +\iota_{\xi}$ leads to
\beq
d_{H}(e^{2A-\Phi} \Psi^{(0)}_{\mp})= \mp \frac{1}{8}c f_{\pm}\, ,
\eeq
from which it follows that in Type IIA $c f_0=0$ and that in general pure R--R sources are only possible when $c=0$.
The vector dual to $\xi$ generates an isometry and in fact is a symmetry of the 
whole background which corresponds to the R-symmetry. Under this symmetry the bi-spinors transform as 
\beq
{\cal L}_{\xi}\Psi^{(0)}= {\cal L}_{\xi}\Psi^{(2)}=0,~~~{\cal L}_{\xi}(\Psi^{(1)}+i \Psi^{(3)})=- 4 i m  (\Psi^{(1)}+i \Psi^{(3)})\, .
\eeq
The background also possesses a symmetry generated by the vector dual of a ten-dimensional 1-form
\beq
K=\frac{1}{32}\big(2e^{2A}k -   \xi\big)\, ,
\eeq
where $k$ is the 1-form dual to a time-like Killing vector on AdS$_3$, see appendix \ref{sec:AdS3app}. Two classes of backgrounds can be dissociated depending on whether $K$ is time-like or null. The latter case occurs for $||\xi|| = 2 e^{A}$.\\
~\\
 All the conditions presented so far hold for $m=0$, for which AdS$_3$ becomes Mink$_3$. Generically $\xi^a\partial_{a}$ is now an isometry with respect to which the spinors and bi-linears are singlets. Clearly there exist $ {\cal N}=2$ Mink$_3$ solutions for which no such isometry exists, the flat space D2-brane with no rotational invariance in its co-dimensions for instance. However for Mink$_3$ it is now possible to fix $\xi=0$ so there is no isometry, this also implies $c=0$ (the converse only implies $\langle \xi,\tilde{\xi}\rangle=0$). This gives a concrete physical interpretation for  $c\neq 0$ in this case, i.e.\ 
\beq
{\cal N}=2~ \text{Mink}_3:~ c\neq 0 ~~~~\Rightarrow~~~~ \text{U(1) flavour isometry}~\xi^a\partial_a\, .
\eeq
It is extremely common that classes of solutions with necessary flavour isometries  can be mapped to more general classes for which this isometry is not necessary after T-duality. Thus we expect the most general classes of ${\cal N}=2$ Mink$_3$ vacua (modulo duality) to be constrained such that $\xi=c=0$. An exploration of such Minkowski vacua is beyond the scope of this work, but would be interesting to pursue. \\
~~\\
In the present work we will focus on AdS$_3$ solutions in Type IIA and take $c=0$, as we are primarily interested in backgrounds with non-zero Romans mass. \\
~\\
In general, the four Majorana spinors $(\chi^1_1,\chi^2_1,\chi^1_2,\chi^2_2)$, can be decomposed in terms of a single unit-norm Majorana spinor $\chi$, and three real unit-norm 1-forms $(V_1, V_2, V_3)$ whose interior products we parameterise as
\beq
\langle V_1, V_2 \rangle = c_3\, ,~~~~\langle V_2, V_3 \rangle = c_1\, ,~~~~\langle V_3, V_1\rangle= c_2\, ,
\eeq
for real functions $(c_1, c_2, c_3)$.  In order to solve
\beq
\chi^{I\dag}_1\chi^J_1=\chi^{I\dag}_2\chi^J_2= e^{A}\delta^{IJ}\, ,
\eeq
we take, without loss of generality, the following parameterisation
\beq
\chi^1_1= e^{\frac{A}{2}}\chi\, ,~~~~
\chi^2_1= -i e^{\frac{A}{2}} V_1\chi\, ,~~~~
\chi^1_2= e^{\frac{A}{2}}(a-i b V_2)\chi\, ,~~~~
\chi^2_2= -ie^{\frac{A}{2}}V_3(a-i b V_2)\chi\, ,
\eeq
where $a, b$ are real functions constrained as 
\beq
a^2+b^2=1 \, .
\eeq
The Majorana spinor $\chi$ defines a $G_2$-structure characterised by a 3-form $\Phi_3$ such that
\beq
\chi\otimes \chi^{\dag}=\Psi^{(G_2)}_++ i \Psi^{(G_2)}_-= \frac{1}{8}(1-i\Phi_3-\star_7 \Phi_3+ i\text{vol}_7)\, ,~~~ \Phi_3\wedge \star_7\Phi_3=7 \text{vol}_7\, .
\eeq
In what follows we will work with  
\beq
\xi_{\pm} \equiv V_1\pm V_3\, ,~~~~ V \equiv V_2\, ,
\eeq
in terms of which
\beq
\xi=-e^{A}\xi_-\, ,~~~~\tilde{\xi}=-e^{A}\xi_+\, .
\eeq
We will also define an auxiliary SU(3)-structure via
\beq
\Phi_3=  \frac{\xi_-}{|\xi_-|}\wedge J_2-\text{Im}\Omega_3\, ,~~~~\star_7\Phi_3= \frac{1}{2}J_2\wedge J_2-\frac{\xi_-}{|\xi_-|}\wedge\text{Re}\Omega_3\, .
\eeq
 A branching of possible solutions now appears depending on how $(V_1,V_2)$ are aligned. Generically both $\xi_{\pm}$ are non trivial and can be used to define components of the vielbein --- the exception is when $V_1=\pm V_3$, which sets one of $\xi_{\pm}$ to zero; these cases need to be analysed separately. For AdS$_3$ we must have $\xi_-\neq 0$, but there is no barrier to fixing $\xi_+=0$, which one can check is equivalent to imposing that the ten-dimensional Killing vector $K^M\partial_M$ is null. The rest of this paper will be focused on classifying such AdS$_3$ solutions and finding new explicit examples.

\section{The null case}\label{sec:null classification}
In this work we will study the case of $K$ being null which is equivalent to $\xi_+ = 0$. From this point we also take
\beq
m\neq 0\, .
\eeq
We are left with $(\xi_-, V)$ and we introduce $v$ such that
\beq
V=\cos\theta v+  \frac{1}{2}\sin\theta\xi_-\, ,~~~~ \iota_v \xi_-=0,~~~~||v||=1\, .
\eeq
We can then further decompose the auxiliary SU(3)-structure as
\beq
J_2= j_2 +u\wedge v\, ,~~~~ \Omega_3=(u+i v)\wedge \omega_2\, ,
\eeq
with $u$ a unit-norm 1-form orthogonal to $(\xi_-, v),$ and $(j_2,\omega_2)$ defining an SU(2)-structure. 

In order to parameterise the $d=7$ bi-spinors in as simple a fashion as possible we find it convenient to decompose the functions of the spinor, and redefine the SU(2)-structure forms as
\beq
a+ i b \sin\theta =\cos\beta e^{i\psi}\, ,~~~~b \cos\theta= \sin\beta\, ,~~~~\omega_2\to e^{-i\psi}\omega_2\, ,~~~z=u+i v  \, ,
\eeq
and introduce 
\beq
\Psi^{(\text{SU}(2))}_+=\frac{e^{i\psi}}{8}\bigg[\cos\beta e^{-i j_2}-\sin\beta\omega_2\bigg]\wedge e^{\frac{1}{2}z\wedge \overline{z}}\, ,~~~\Psi^{(\text{SU}(2))}_-=\frac{1}{8}z \wedge\bigg[\sin\beta e^{-i j_2}+\cos\beta\omega_2\bigg]\, .
\eeq
In terms of these we have
\begin{align}
\Psi^{(0)}_+&=\xi_-\wedge \text{Re}\Psi^{(\text{SU}(2))}_-\, ,~~~~&\Psi^{(0)}_-&=2\text{Im}\Psi^{(\text{SU}(2))}_-\, , \nn \\[2mm]
\Psi^{(1)}_+&=2\text{Im}\Psi^{(\text{SU}(2))}_+\, ,~~~~&\Psi^{(1)}_-&=-\xi_-\wedge \text{Re}\Psi^{(\text{SU}(2))}_+\, , \nn \\[2mm]
\Psi^{(2)}_+&=-\xi_-\wedge \text{Im}\Psi^{(\text{SU}(2))}_-\, ,~~~~&\Psi^{(2)}_-&=2  \text{Re}\Psi^{(\text{SU}(2))}_-\, , \nn \\[2mm]
\Psi^{(3)}_+&=2\text{Re}\Psi^{(\text{SU}(2))}_+\, ,~~~~&\Psi^{(3)}_-&=\xi_-\wedge\text{Im}\Psi^{(\text{SU}(2))}_+\, .
\end{align}
One can readily check that the supersymmetry equations impose 
\beq
\cos\beta = 0\, ,
\eeq
without loss of generality. In this case, the first non-trivial component of \eqref{eq:bps2} is the 3-form, which imposes
\beq
d(e^{3A-\Phi}e^{i\psi}\omega_2)+i e^{2A-\Phi}m e^{i\psi}\xi_-\wedge \omega_2=0\,.
\eeq
Taking the general ansatz
\beq
\frac{1}{2}\xi_-= e^{C}(d\psi+ {\cal A}) \, ,
\eeq
then fixes
\beq
 e^{C}=-\frac{e^{A}}{2m}\,,~~~~ 
 d(e^{3A-\Phi}\omega_2)=i {\cal A}\wedge (e^{3A-\Phi}\omega_2)\,.
\eeq
The 5-form component of \eqref{eq:bps2} then yields
\beq
\big(H+i d(u\wedge v)\big)\wedge\omega_2=0\, .
\eeq
Equation \eqref{eq:bps1} contains the following constraints:
\begin{align}
d(e^{2A-\Phi}v)&=0\, ,\\[2mm]
d(e^{2A-\Phi}u\wedge j_2)+e^{2A-\Phi} H\wedge v&=0\, ,\\[2mm]
(dj_2\wedge v-  H\wedge u )\wedge j_2&=0\, .
\end{align}
We may decompose the flux $f_+$ as
\beq
f_+= g_++ \frac{1}{2}\xi_-\wedge g_-\label{eq:magfluxesgen}\, ,
\eeq
and given what has been derived thus far, \eqref{eq:bps4} imposes
\beq\label{eq:gminus}
g_-=  -\frac{e^{A-\Phi}}{2m}(v-u\wedge j_2)\wedge {\cal F}\, ,
\eeq
where ${\cal F} = d {\cal A}$.
Now since $\Psi^{(0)}_-$ is orthogonal to $\xi_-$ we immediately get from  \eqref{eq:bps5} that
\beq
({\cal F}- 2m^2 e^{-2A} j_2)\wedge j_2\wedge v\wedge u=0\, .\label{eq:forRicci}
\eeq
Combining this with $ {\cal F}\wedge \omega_2=0$ implies 
\beq\label{eq:calfdef0}
{\cal F}=2m^2 e^{-2A} (j_2+{\cal F}^{(1,1)})\, ,
\eeq
where ${\cal F}^{(1,1)}$ is a primitive $(1,1)$-form.

To proceed it is helpful to consider the torsion classes of an SU(2)-structure in six dimensions, see  appendix \ref{app:torsions} for the general form of these. We shall compute the torsion classes on 
\beq
ds^2(\hat{\text{M}}_6)= e^{3A-\Phi}ds^2(\text{M}_6)\,,
\eeq
where we add hats to the various forms to indicate this. The relevant classes are
\beq
d\hat{u}=s_1\text{Re}\hat\omega_2+s_2\text{Im}\hat\omega_2+s_3 \hat j_2+s_4\hat u\wedge \hat v+T_1+ \hat u\wedge W_1+\hat v\wedge W_2\,,~~~~~ d\hat j_2= W_3\wedge \hat {j}_2+ \hat u\wedge T_2+\hat v\wedge T_3\,,
\eeq
with $s_i$ real functions, $W_i$ real 1-forms and $T_i$ real primitive (1,1)-forms. We will also introduce some holomorphic 1-forms along the way,  $V_i$. Expanding the NS--NS field-strength as
\beq
H=H_3+ \hat u\wedge H^1_2+\hat v\wedge H^2_2+\hat u\wedge \hat  v \wedge H_1\, ,
\eeq
and the exterior derivative as\footnote{The tilde on $d_4$ indicates a potentially twisted derivative as $u$ may be fibered over M$_4$, $v$ can only be trivially fibred.}
\beq
d= \tilde{d}_4+  \hat u \wedge \iota_{\hat u} d+  \hat v \wedge \iota_{\hat v} d\, ,
\eeq
then plugging all this into the derived constraints we find they reduce to 
\begin{align}
&d(e^{\frac{1}{2}(A-\Phi)}\hat v)=0\, ,~~~~ d\hat \omega_2= i {\cal A}\wedge \hat \omega_2\, ,~~~~e^{3A-\Phi}H_1=\text{Re}V_1\, ,~~~~~\text{Im}V_1=-W_1-\frac{1}{2}\tilde{d}_4(7A-3\Phi)\, , \nn\\[2mm]
&s_3=0\, ,~~~~W_3= W_1+\frac{1}{2}\tilde{d}_4(5A-\Phi)\, ,~~~~e^{3A-\Phi}H_3= W_2\wedge \hat j_2\, ,\nn\\[2mm]
&e^{3A-\Phi}H^1_2= T_3-\frac{1}{2}(\iota_{\hat{v}} d(5A-\Phi)+2 s_4)\hat j_2\, ,\nn\\[2mm]
&e^{3A-\Phi}H^2_2=H^{(1,1)}+(s_2 \text{Re}\hat\omega_2-s_1 \text{Im}\hat\omega_2)+ \iota_{\hat{u}}d(3A-\Phi)\hat j_2\, .
\end{align}
For the remaining flux component $g_+$, from \eqref{eq:bps3} we have
\begin{align}
\star_7 \lambda(g_+)&=e^{C-\Phi}D\psi\wedge\bigg[v\wedge(W_1\wedge j_2-W_2)+\frac{1}{2}v\wedge \tilde{d}_4(3A+\Phi)\wedge j_2-u\wedge (W_1+W_2\wedge j_2)+\frac{1}{2}u\wedge  \tilde{d}_4(5A-\Phi)\nn\\[2mm]
-&e^{-\frac{1}{2}(3A-\Phi)}T_1-e^{\frac{1}{2}(3A-\Phi)}(s_1\text{Re}\omega_2+s_2\text{Im}\omega_2)+e^{\frac{1}{2}(3A-\Phi)}u\wedge v\wedge \bigg(\frac{1}{2}(\iota_{\hat{v}}d(5A-\Phi)-2 s_4)\nn\\[2mm]
-&2\iota_{\hat{u}}Aj_2-\me^{\Phi-3A}H^{(1,1)}-e^{\Phi-3A}T_2-(s_2\text{Re}\omega_2-s_1\text{Im}\omega_2)-\frac{1}{4}(\iota_{\hat{v}}(3 A+\Phi)+2s_4)j_2\wedge j_2\bigg)\bigg]\, ,
\end{align}
where the lack of hats is intentional, we need to take the Hodge dual of this after all. To do so we define
\beq
\text{Re}V_2= W_1\, ,~~~~\text{Re}V_3= W_2\, ,~~~~\text{Re}V_4= \tilde{d}_4A\, ,~~~~\text{Re}V_5= \tilde{d}_4\Phi\, ,
\eeq
 and we find
\begin{align}
f_0=g_0&=-\frac{1}{2}e^{\frac{3}{2}(A-\Phi)}(\iota_{\hat{v}}d(3A+\Phi)+2 s_4)\, ,~~~~~~g_6=0\, ,\nn\\[2mm]
g_2&=e^{\frac{3}{2}(A-\Phi)}\bigg[2\iota_{\hat u}\dd Aj_2+s_2\text{Re}\omega_2-s_1 \text{Im}\omega_2-e^{\Phi-3A}T_2-e^{\Phi-3A}H^{(1,1)}\nn\\[2mm]
&+ e^{-\frac{1}{2}(3A-\Phi)}\text{Im}V_3\wedge v+e^{-\frac{1}{2}(3A-\Phi)}\text{Im}(V_2+\frac{3}{2}V_4+\frac{1}{2}V_5)\wedge u\bigg]\, ,\nn\\[2mm]
g_4&=e^{\frac{3}{2}(A-\Phi)}\bigg[\frac{1}{4}(\iota_{\hat v}d(5A-\Phi)-2 s_4)j_2\wedge j_2-u\wedge v\wedge (s_1 \text{Re}\omega_2+s_2\text{Im}\omega_2-e^{-3A+\Phi}T_1)\nn\\[2mm]
&+e^{-\frac{1}{2}(3A-\Phi)}\text{Im}V_3\wedge j_2\wedge u-e^{-\frac{1}{2}(3A-\Phi)}\text{Im}(V_2-\frac{5}{2}V_4+\frac{1}{2}V_5)\wedge j_2\wedge v\bigg]\, .\label{eq:gs}
\end{align}

\subsection{Introducing coordinates}

Above we have presented the general conditions for a solution to preserve supersymmetry. In this section we will further reduce the system of equations by introducing coordinates for the system. Firstly, since $\hat{v}$ is conformally closed we can introduce a coordinate via
\begin{equation}
\hat{v}=\me^{\tfrac{1}{2}(\Phi-A)} \dd y\, .
\end{equation}
We may introduce an additional coordinate for $\hat{u}$ via
\begin{equation}
\hat{u}=\me^{U}(\dd \varphi+ \sigma+ \tau \dd y)\equiv \me^U D\varphi\, ,
\end{equation}
where $\sigma$ has legs only along the SU$(2)$-structure manifold. It is natural to assume that both $\sigma$ and $\tau$ are independent of $\varphi$, and moreover it is natural to assume $\partial_\varphi$ is a Killing direction. For the time being we will not assume this, but instead reduce to this more restrictive class in section \ref{sec:Additional symmetries} and construct explicit solutions there. 

With these coordinates we have,
\begin{equation}
i_{\hat{u}}\dd =e^{-U}\partial_\varphi\, ,\qquad i_{\hat{v}}\dd= \me^{\tfrac{1}{2}(A-\Phi)}\Big(\partial_y-\tau\partial_\varphi\Big)\equiv \me^{\tfrac{1}{2}(A-\Phi)}\tilde{\partial}_y\, ,
\end{equation}
and the exterior derivative takes the form
\begin{equation}
\dd= \dd y \wedge \tilde{\partial}_y+D\varphi\wedge \partial_\varphi +\tilde{\dd}_4\, ,
\end{equation}
where
\be
\tilde{\dd}_4=\dd_4 -\sigma\partial_\varphi\, ,\qquad \tilde{\partial}_y=\partial_y -\tau \partial_\varphi\, .
\ee
Note that the twisted exterior derivative $\tilde{\dd}_4$ satisfies
\be
\tilde{\dd}_4^2=-(\tilde{\dd}_4 \sigma)\wedge \partial_\varphi\, ,
\ee
and is therefore generically not nilpotent. Note, that for this to define a genuine exterior derivative, as opposed to twisted, we require $s_1=s_2=T_1=0$ or for $\partial_\varphi$ to be a Killing vector.

Using the local coordinates we may further decompose the torsion conditions. 
For $\hat{u}$ we find
\begin{align}
\tilde{\dd}_4 \sigma&= \me^{-U}(s_1 \Real \hat{\omega}_2+s_2\Imag\hat{\omega}_2+T_1)\, ,\\
\partial_\varphi\sigma-\tilde{\dd}_4U&=W_1\, ,\\
\partial_\varphi \tau-\tilde{\partial}_yU&=s_4 \me^{\tfrac{1}{2}(\Phi-A)}\,\label{eq:duy}\, , \\
\me^{U-\tfrac{1}{2}(\Phi-A)}(\tilde{\partial}_y \sigma-\tilde{\dd}_4 \tau)&=W_2\,.
\end{align}
A similar decomposition for $\hat{j}_2$ gives
\begin{align}
\dd \hat{j}_2&= \frac{1}{2}\Big(2\partial_\varphi \sigma-\tilde{\dd}_4(2U -5A+\Phi)\Big)\wedge \hat{j}_2+\me^U D\varphi\wedge T_2 +\me^{\tfrac{1}{2}(\Phi-A)}\dd y \wedge T_3\, ,
\end{align}
which in components reads
\begin{align}
\tilde{\dd}_4\hat{j}_2&=\frac{1}{2}\Big(2\partial_\varphi\sigma-\tilde{\dd}_4(2U-5 A+\Phi)\Big)\wedge \hat{j}_2\, ,\\
\partial_\varphi\hat{j}_2&=\me^U T_2\, ,\\
\tilde{\partial}_y\hat{j}_2&=\me^{\tfrac{1}{2}(\Phi-A)}T_3\, .
\end{align}
Observe that if $\varphi$ defines a symmetry, then $\hat{j}_2$ is conformally closed.

Next, consider the decomposition of the connection one-form $\mathcal{A}$, 
\be
\mathcal{A}=P+\mathcal{A}_y\dd y+\mathcal{A}_\varphi D\varphi\, .
\ee
With this decomposition we may decompose the torsion conditions for $\hat{\omega}_{2}$, which gives
\begin{align}
\tilde{\dd}_4 \hat{\omega}_2&=\ii P\wedge \hat{\omega}_2\, ,\\
\tilde{\partial}_y\hat{\omega}&=\ii \mathcal{A}_y\hat{\omega}_2\, ,\\
\partial_\varphi\hat{\omega}&=\ii \mathcal{A}_\varphi \hat{\omega}_2\,,
\end{align}
where the first of these implies that the four-manifold supporting the SU(2)-structure forms is complex, with an associated complex structure $I$. 
Finally the torsion condition 
\eqref{eq:forRicci}
leads to 
\be
\Big(\tilde{\dd}_4 P-2 m^2 \me^{\Phi-5 A}\hat{j}_2\Big)\wedge \hat{j}_2=0\, .\label{eq:ForRicci}
\ee

Having rewritten the torsion conditions in terms of the local coordinates we can proceed with the decomposition of the fluxes in coordinate form. The NS--NS 3-form field-strength is 
\begin{align}
H&=\me^{U+\tfrac{1}{2}(\Phi-5A)} \Big(\tilde{\partial}_y \sigma-\tilde{\dd_4}\tau\Big)\wedge \hat{j}_2+\me^{U+\Phi-3A}D\varphi\wedge \bigg(T_3 -\frac{1}{2}\me^{\tfrac{1}{2}(A-\Phi)}\Big(2 \partial_\varphi \tau-\tilde{\partial}_{y}(2U+\Phi-5A)\Big)\hat{j}_2\bigg)\nonumber\\
&+\me^{\tfrac{1}{2}(3\Phi-7A)}\dd y \wedge \bigg( H^{(1,1)}+s_2\Real\hat{\omega}_2-s_1\Imag\hat{\omega}_2+\me^{-U}\partial_\varphi(3 A-\Phi)\hat{j}_2\bigg)\nonumber\\
&+ \me^{U+\tfrac{1}{2}(3\Phi-7A)}D \varphi\wedge \dd y \wedge \Big(-I\cdot (\partial_\varphi \sigma)-\frac{1}{2} \tilde{\dd}_4^c(2 U+3\Phi-7A)\Big)\,,
\end{align}
with 
\be
\tilde{\dd}_4^{c}\equiv -I\cdot (\tilde{\dd}_4)\, .
\ee

We may rewrite the condition for the Romans mass as
\be
\frac{1}{2}\tilde{\partial}_y (2U-\Phi-3A)-\me^{2(\Phi-A)}f_0=\partial_\varphi \tau\, ,
\ee
by using \eqref{eq:duy}. The 2-form $g_2$ takes the form
\begin{align}
g_2&=\me^{-\tfrac{1}{2}(\Phi+3A)}\Big[ s_2\Real \hat{\omega}_2-s_1 \Imag \hat{\omega}_2-T_2-H^{(1,1)}\Big]-\me^{U-\tfrac{1}{2}(\Phi+3A)}I\cdot \Big(\tilde{\partial}_y \sigma -\tilde{\dd}_4 \tau\Big)\wedge \dd y \nonumber\\
&+\me^{U -\tfrac{1}{2}(\Phi+3A)}I\cdot \Big(\tilde{\dd}_4\big(U-\tfrac{1}{2}(3 A+\Phi)\big)-\partial_\varphi\sigma\Big)\wedge D\varphi+2\me^{U-\tfrac{1}{2}(\Phi+3A)}\partial_\varphi A~ \hat{j}_2\, ,
\end{align}
whilst the 4-form $g_4$ is
\begin{align}
g_4&=\me^{-4 A} \Big( \tilde{\partial}_y \big(U+\tfrac{1}{2}(5 A-\Phi)\big)-\partial_\varphi \tau\Big)\frac{1}{2}\hat{j}_2\wedge\hat{j}_2-\me^{U+\Phi-5 A}D\varphi\wedge \dd y\wedge \Big( s_1\Real\hat{\omega}_2+s_2 \Imag\hat{\omega}_2-T_1\Big)\nonumber\\
&-\me^{2 U-4A}I \cdot\Big(\tilde{\partial}_y\sigma -\tilde{\dd}_4\tau\Big)\wedge \hat{j}_2\wedge D\varphi+\me^{\Phi-5A}I\cdot \Big(\partial_\varphi\sigma -\tilde{\dd}_4 \big(U +\tfrac{1}{2}(5A-\Phi)\big)\Big)\wedge \hat{j}_2\wedge \dd y\, .
\end{align}
Finally the Bianchi identity reads
\be
\dd g_+-H\wedge g_+=- \frac{1}{4 m^2} \dd y\wedge \dd \mathcal{A}\wedge \dd \mathcal{A}\, .
\ee

\section{Solutions with additional symmetries}\label{sec:Additional symmetries}

In the above we have presented the general decomposition of the torsion conditions using a set of coordinates. The resultant conditions are difficult to solve and therefore to make further progress we will impose some assumptions which make the problem more tractable. 

\subsection{Imposing an additional Killing vector}

As we emphasised earlier one natural assumption to make is to impose that $\partial_\varphi$ is a Killing vector. We will therefore assume that all the scalars and 2-form $\hat{j}_2$ are independent of $\varphi$. This assumption lets us drop $\tilde{\dd}_4$ for $\dd_4$ since everything (but $\hat{\omega}_2$) is independent of $\varphi$. It follows that $\hat{j}_2$ is conformally K\"ahler
\be
\dd_4\Big(\me^{U+\tfrac{1}{2}(\Phi-5A)} \hat{j}_2\Big)=0\, ,
\ee
and we therefore redefine our K\"ahler form to be
\be
J= \me^{\tfrac{1}{2}(2U-5A+\Phi)} \hat{j}_2\, . 
\ee
This implies that the internal metric takes the form
\be
\dd s_7^2=\frac{\me^{2A}}{4m^2}(\dd \psi +\mathcal{A})^2+\me^{\Phi-3 A}\bigg[\me^{2 U}D\varphi^2+\me^{\Phi-A}\dd y^2+\me^{\tfrac{1}{2}(5A-\Phi)- U} \dd s^2(\text{M}_4)\bigg]\,,
\ee
with the metric on M$_4$ K\"ahler at fixed $y$ coordinate.

The torsion conditions for the new K\"ahler 2-form read
\begin{align}
\dd_4J&=0\, , \label{eq:Kahlercond}\\
\partial_ \varphi J&=\me^{\tfrac{1}{2}(4U-5A+\Phi)} T_2\, ,\\
\partial_y J&=\frac{1}{2}\partial_y (2U-5A+\Phi) J +\me^{U+\Phi-3A} T_3\, ,\label{eq:dyJ}
\end{align}
however since we assume $\varphi$ is an isometry we must set $T_2=0$ in the following.

Performing the same rescaling for the holomorphic volume form,
\be
\Omega=\me^{\tfrac{1}{2}(2U-5A+\Phi)}\hat{\omega}_2\, ,
\ee
we find the torsion conditions
\begin{align}
\dd_4 \Omega&=\Big(\ii (P+\mathcal{A}_\varphi \sigma)+\frac{1}{2}\dd_4(2U-5A+\Phi)\Big)\wedge \Omega\, ,\nonumber\\
\partial_y \Omega&=\Big(\ii (\mathcal{A}_y +\mathcal{A}_\varphi\tau)+\frac{1}{2}\partial_y(2U-5A+\Phi)\Big)\Omega\, ,\label{eq:d4Gammasimp}\\
\partial_\varphi\Omega&=\ii \mathcal{A}_\varphi \Omega\, .\nonumber
\end{align}
It follows that $\mathcal{A}_\varphi$ should be a constant and therefore we may solve the final constraint simply by introducing a phase for $\Omega$. Moreover, integrability implies 
\begin{align}
\partial_y P-\dd_4 \mathcal{A}_y +\mathcal{A}_\varphi(\partial_y \sigma -\dd_4 \tau)&=0\, ,\\
\dd_4 P+\mathcal{A}_\varphi \dd_4\sigma &\in H^{(1,1)}(M_4)\, .
\end{align}
For non-trivial $\mathcal{A}_\varphi$ we may solve for $P$ in terms of $\sigma$ up to the addition of a term whose derivative is a $(1,1)$-form. In addition we have that 
\be
\dd_4\sigma= \me^{-U}\Big(s_1 \me^{-\tfrac{1}{2}(2U -5 A+\Phi)}\Real \Omega+s_2\me^{-\tfrac{1}{2}(2U -5 A+\Phi)} \Imag \Omega +T_1\Big)\, .\label{eq:d4sigma}
\ee
It is useful to redefine $s_1$ and $s_2$ here to absorb the exponential factors but since the classes of solutions we consider later do not have such a term switched on we will refrain from doing so here. The supersymmetry condition \eqref{eq:ForRicci} becomes
 \be
 \Big(\dd_4 P-2 m^2 \me^{- U -\tfrac{1}{2}(5 A -\Phi)}J\Big)\wedge J=0\, ,
 \ee
 which we may rewrite as the scalar equation
 \be
 R_4=2\square_4\Big(U-\tfrac{1}{2}(5A-\Phi)\Big)+8 m^2 \me^{-U -\tfrac{1}{2}(5 A -\Phi)}\, .\label{eq:Riccifix}
 \ee
 Note that if the first term on the right-hand side vanishes this is reminiscent of the condition for GK geometries \cite{Kim:2005ez,Kim:2006qu,Gauntlett:2007ts} after a little redefinition.\footnote{In \cite{Couzens:2020jgx} similar types of condition for the Ricci scalar appears with a Laplacian like term. The geometries of \cite{Couzens:2020jgx} generalise GK geometries by including rotation, though this is not the origin of such a term here.} As we will see later one of the classes of solution we obtain are the T-dual of the GK geometries in Type IIB with a torus and 3-form flux in massless Type IIA. In fact we are able to generalise these solutions further by turning on a non-trivial Romans mass.
 
 The NS--NS 3-form after this simplification becomes
 \begin{align}
 H&=\Big(\partial_y\sigma-\dd_4\tau\Big)\wedge J+D\varphi\wedge \Big(\me^{U +\Phi-3 A}T_3 +\frac{1}{2}\partial_y (2 U +\Phi-5 A)J\Big)\label{eq:NSsimp}\\
 &+\me^{-U -A+\Phi}\dd y\wedge \Big(\me^{U -\tfrac{1}{2} (5A-\Phi)}H^{(1,1)}+s_2 \Real\Omega-s_1 \Imag\Omega\Big)-D\varphi\wedge \dd y\wedge \dd_4^c\Big(\me^{U +\tfrac{1}{2}(3\Phi-7A)}\Big)\, .\nonumber
 \end{align}
The condition for the Romans mass simplifies to 
\begin{align}
\partial_y(2U-\Phi-3A)=2\me^{2(\Phi-A)}f_0\, ,\label{eq:Rmasseq}
\end{align}
whilst the R--R 2-form becomes
\begin{align}
g_2&=\me^{-U-\Phi+A}\Big(s_2 \Real\Omega -s_1 \Imag \Omega-\me^{U-\tfrac{1}{2}(5 A-\Phi)}H^{(1,1)}\Big)-\me^{U-\tfrac{1}{2}(\Phi+3 A)}I\cdot (\partial_y \sigma -\dd_4 \tau)\wedge \dd y \nonumber\\
&-\dd_4^c\Big( \me^{U-\tfrac{1}{2}(\Phi+3A)}\Big)\wedge D\varphi\, ,\label{eq:g2simp}
\end{align}
and the 4-form is
\begin{align}
g_4&=\me^{-2 U +A -\Phi} \partial_y \big(U+\tfrac{1}{2}(5A-\Phi)\big)\frac{1}{2}J\wedge J -D\varphi\wedge \dd y \wedge \Big(\me^{\tfrac{1}{2}(\Phi-5A)}(s_1 \Real \Omega +s_2 \Imag \Omega) -\me^{U +\Phi-5A}T_1\Big)\nonumber\\
&-\me^{U -\tfrac{1}{2}(\Phi+3A)}I\cdot (\partial_y \sigma -\dd_4 \tau)\wedge J\wedge D\varphi -\dd_4^c \big(\me^{- U-\tfrac{1}{2}(5A-\Phi)}\big)\wedge J \wedge \dd y\, .\label{eq:g4simp}
\end{align}

We may construct the magnetic fluxes using \eqref{eq:magfluxesgen} and find 
\begin{align}
f_2&= g_2\, ,\nonumber\\
f_4&=g_4+\frac{1}{4m^2}(\dd\psi+\mathcal{A})\wedge \dd y\wedge \dd\mathcal{A}\, ,\\
f_6&=-\frac{1}{4m^2}(\dd \psi+\mathcal{A})\wedge D\varphi\wedge J\wedge \dd\mathcal{A}\, .
\end{align}

\subsection{Product of Riemann surface ansatz}

Having introduced coordinates and made the assumption of an additional Killing vector we are in a position where we can introduce an ansatz for the four-dimensional base. The simplest choice is that we may decompose the base as the direct product of two Riemann surfaces. We take
\be
\dd s^2(M_4)=\me^{2 f_1(y)}\dd s^2(\Sigma_1)+\me^{2 f_2(y)}\dd s^2(\Sigma_2)\, ,
\ee
with $f_i(y)$ arbitrary functions of $y$ and we take
the metric on $\Sigma_{i}$ to be the constant curvature one given by 
\be
\dd s^2(\Sigma_i)=\frac{1}{1-\kappa_i x_i^2}\dd x_i^2+\big(1-\kappa_i x_i^2\big)\dd z_{i}^2\, .
\ee
The Ricci scalar for the Riemann surface is $R=2\kappa_i$ and we take the structure forms to be
\be
J_i=\dd x_i\wedge \dd z_i\, ,\qquad \Omega_i=\frac{1}{\sqrt{1-\kappa_i x_i^2}}\Big(\dd x_i+\ii (1-\kappa_{i}x_i^2)\dd z_i\Big)\, ,
\ee
which satisfy
\be
\dd_{4} J_i=0\, ,\quad \dd_4 \Omega_i=\ii \kappa_i x_i \dd z_i\wedge \Omega_i\, .
\ee
In terms of the structure forms of the Riemann surfaces the SU$(2)$-structure forms are
\be
J=\me^{2 f_1(y)}J_1+\me^{2 f_2(y)}J_2\, ,\quad \Omega=\me^{f_1(y)+f_2(y)}\Omega_1\wedge \Omega_2\, .
\ee
For the given K\"ahler form we can construct a single primitive $(1,1)$-form which preserves the symmetries, namely
\be
\prim=\me^{2 f_1(y)}J_1-\me^{2 f_2(y)}J_2\, .\label{eq:nudef}
\ee
Note that $\dd_4 \prim=0$, indeed without breaking the symmetries of the Riemann surfaces no other choice is possible. 

Let us define
\be
T_3= t'_3(y)\me^{-U(y)-\Phi(y)+3 A(y)}\prim\, ,
\ee
then \eqref{eq:dyJ} implies
\begin{align}
4f_1(y)&=2U(y)-5 A(y) +\Phi(y)+2 t_3(y)+c_1\, ,\nonumber\\
4f_2(y)&=2U(y)-5 A(y) +\Phi(y)-2 t_3(y)+c_2\, .
\end{align}
We solve for $t_3(y)$ and $U(y)$, giving
\begin{align}
\me^{2U(y)}&=C_1^2\me^{2 f_1(y)+2f_2(y)+5 A(y)-\Phi(y)}\, ,\label{eq:Usol}\\
t_3(y)&=C_2 +f_1(y)+f_2(y)\, .\label{eq:t1sol}
\end{align}
Next take the primitive 2-form $T_1$ to be
\be
T_1=\me^{U(y)}t_1(y)\prim\, ,
\ee
the integrability condition for \eqref{eq:d4sigma} implies $s_1=s_2=0$ unless $\kappa_1=\kappa_2=0$. Let
\be
A_i= x_i\dd z_i\, ,
\ee
then $\sigma$ takes the form
\be
\sigma= \sum_{i=1}^{2}u_i(y) A_i\, .
\ee
Plugging this into \eqref{eq:d4sigma} gives
\be
\dd_4\sigma=\sum_{i=1}^{2} u_i(y)  J_i= t_1(y)\prim\, ,
\ee
and therefore
\be
u_1(y)=t_1(y)\me^{2 f_1(y)}\, ,\quad u_2(y)=-t_1(y)\me^{2 f_2(y)}\, .
\ee
It follows that a non-trivial $T_1$ leads to $D\varphi$ being non-trivially fibered over the base. Next consider the conditions on the holomorphic volume form \eqref{eq:d4Gammasimp}. From the first we find
\be
P+\mathcal{A}_\varphi \sigma=\kappa_1 A_1+\kappa_2 A_2\, ,
\ee
whilst the second implies 
\be
\mathcal{A}_y=-\mathcal{A}_\varphi \tau\, .
\ee
We must require that $\tau$ is a function of $y$ only, and does not have any Riemann surface dependence. It then follows that a coordinate transformation can be performed which sets $\tau=0$ and therefore without loss of generality we may take $\tau=0$ and therefore also $\mathcal{A}_y=0$.
 Note that $\mathcal{A}_\varphi$ is a constant which we can pick by rescaling the holomorphic volume form by a $\varphi$ dependent phase, we can therefore set it to vanish without loss of generality. It follows that the 1-form $P$ is
\be
P=\kappa_1 A_1+\kappa_2 A_2\, .
\ee

From the expression for the NS--NS flux in \eqref{eq:NSsimp} the Bianchi identity imposes
\be
\partial_y\Big(\me^{2 f_1(y)}f_1'(y)\Big)=\partial_y\Big(\me^{2 f_2(y)}f_2'(y)\Big)=\partial_{y}^2\Big(\me^{2 (f_1(y)+f_2(y))} t_1(y)\Big)=0\, ,
\ee
note that it is independent of the primitive two-form $H^{(1,1)}$ which is necessarily closed on the four-dimensional base in order to preserve the symmetries of the Riemann surfaces. 
We may solve the first two by 
\be
\me^{2f_i(y)}= a_i y+b_i\, ,
\ee
which, upon substituting into the third gives 
\be
t_1(y)=\frac{ \alpha y+\beta}{(a_1 y+b_1)(a_2 y+b_2)}\, .
\ee
So far we have solved for $t_1(y), t_3(y), f_i(y), U(y)$ and it remains to determine $A(y)$ and $\Phi(y)$. From \eqref{eq:Riccifix} we find
\be
\me^{-\Phi}= \frac{4 m^2 \me^{f_1(y)+f_2(y)-5 A(y)}}{C_1 (\kappa_1 \me^{2 f_2(y)}+\kappa_2 \me^{2 f_1(y)})}\, .\label{eq:Phisol1}
\ee
The condition from the Romans mass reads
\be
\partial_y \Big(2 U(y)-\Phi(y)-3A(y)\Big)=2 \me^{2\big(\Phi(y)-A(y)\big)} f_0\, ,
\ee
which we may solve for $A$ giving
\begin{align}
\me^{8A(y)}=\frac{48 m^2 (a_1 y+b_1)^2(a_2 y+b_2)^2}{C_1^2 f_0(\kappa_1(a_2 y+b_2)+\kappa_2 (a_1 y+b_1))^2 \big(3 b_1 y(2 b_2+a_2 y)+a_1 y^2(3 b_2+2 a_2 y)-12 \delta\big)}\, ,
\end{align}
in the massive case and
\be
\me^{4A(y)}=\frac{\hat{\delta} (b_1+a_1 y)(b_2+a_2 y)}{\kappa_1 (b_2+a_2 y)+\kappa_2(b_1+a_1 y)}\, ,
\ee
for the massless case.\footnote{Note that if one redefines the constant $\delta\rightarrow -4 m^4(f_0C_1^2\hat{\delta}^2)^{-1}$ in the massive case one can take the massless limit and land upon the massless solution we have presented.}

We have now solved for all the functions appearing in the solution, but for the primitive two-form $H^{(1,1)}$, using the Bianchi identity for the NS--NS flux, the Romans mass Bianchi identity and the Ricci scalar supersymmetry equation. In solving these conditions we have introduced eight integration constants, $(C_1, a_1,a_2,b_1,b_2,\alpha,\beta,\delta)$. We will see that the two remaining Bianchi identities will restrict these integration constants further. Recall that the primitive form $H^{(1,1)}$ was not constrained by the Bianchi identity for $H$, a convenient choice to make is
\be
H^{(1,1)}=\me^{-\tfrac{1}{2}(\Phi(y)-7 A(y))}h^{(1,1)}(y)\prim\, ,
\ee
with $\prim$ as defined in \eqref{eq:nudef}. This is the most general form we can pick without breaking additional symmetries of the Riemann surfaces. 

The Bianchi identity for $g_2$ then imposes 
\begin{align}
0&=a_{1} f_0=a_2 f_0=\alpha f_0\, ,\nonumber\\
h^{(1,1)}(y)&=\frac{q_1}{(a_1 y+b_1)\sqrt{6 b_1 b_2 y+ 3 (a_2 b_1 +a_1 b_2)y^2 +2 a_1 a_2 y^3-12\delta}}\, ,\nonumber\\
q_2 (a_1 y+b_1)&=q_1 (a_2 y+b_2)\, ,
\end{align}
where $q_2$ is some constant of proportionality.
It is clear to see from the above conditions that there are different branches of solutions to consider. The Bianchi identity for $g_4$ leads to further branching conditions and it is therefore convenient to first solve the $g_2$ Bianchi identity before attempting to solve the $g_4$ one. We will first consider the massive case which turns out to have a unique family of solutions, before studying the massless case.

\subsubsection{Massive class}

We see from above that we must set 
\be
a_1=a_2=\alpha=0\, ,
\ee
which implies that $h^{(1,1)}(y)$ is equal to
\be
h^{(1,1)}(y)=\frac{q}{\sqrt{2 b_1 b_2 y-4\delta}}\,.
\ee
The $g_2$ Bianchi identity is completely solved with these restrictions and we may move onto the $g_4$ Bianchi identity. One finds that this Bianchi identity is solved if we set $q=0$ and thus the primitive two-form $H^{(1,1)}=0$ and 
\be
\beta=\frac{\mathcal{A}_{\varphi}(b_2\kappa_1-b_1 \kappa_2)\pm \sqrt{\mathcal{A}_{\varphi}^2(b_2 \kappa_1+b_1\kappa_2)^2+16b_1^2 b_2^2 C_1^2 m^2 }}{2(\mathcal{A}_{\varphi}^2+4 b_1 b_2 C_1^2 m^2)}\, .
\ee
It is useful to make the redefinitions 
\begin{align}
L f_0=\hat{f}_0
\, ,\quad  y=\frac{2\delta}{b_1b_2}+\frac{L h_8(\hat{y})^2}{2 \hat{f_0}}\, ,\quad h_8(\hat{y})=\hat{f}_0\hat{y}+\hat{c}\, ,
\end{align}
so that 
\be
f_0=\partial_{\hat{y}}h_8(\hat{y})\, ,
\ee
and for simplicity to set $\mathcal{A}_{\varphi}=0$ by multiplying the holomorphic volume form by a suitable $\varphi$-dependent phase. Let us also define
\begin{align}
L&=m^{-1}\, ,\quad b_i=L^2 \hat{b}_i\, ,\quad C_1=L^{-2}\hat{C}_1\, ,\quad   \varphi=L \hat{\varphi}\, ,\nonumber\\
\me^{B}&=\frac{\hat{C}_1(\hat{b}_2\kappa_1+\hat{b}_1\kappa_2)}{4 \sqrt{\hat{b}_1 \hat{b}_2}}\, ,\quad \hat{\beta}=\frac{\sqrt{\kappa_1\kappa_2}}{2 }\, .
\end{align}

The final metric is
\begin{align}
\dd s^2&=L^{2}\frac{\me^{-B/2}}{\sqrt{h_8(\hat{y})}}\bigg[\dd s^2(\text{AdS}_3)^{\text{unit}}+\frac{1}{4}(\dd\psi+\kappa_1 A_1+\kappa_2 A_2)^2+\hat{C}_1^2\hat{b}_1 \hat{b}_2 \Big(\dd\hat{\varphi}+\frac{\hat{\beta}}{\hat{C}_1\hat{b}_1 \hat{b}_2}(\hat{b}_1 A_1-\hat{b}_2 A_2)\Big)^2\nonumber\\
&+\me^{B}\bigg(h_8(\hat{y})\dd \hat{y}^2+\frac{1}{\hat{C}_1\sqrt{\hat{b}_1\hat{b}_2}}\big(\hat{b}_1\dd s^2(\Sigma_1)+\hat{b}_2\dd s^2 (\Sigma_2)\big)\bigg)\bigg]\, ,
\end{align}
with dilaton and magnetic fluxes
\begin{align}
\me^{-4\Phi}&=\me^{B} h_8(\hat{y})^5\, ,\quad H=0\, , \quad 
f_2=0\, , \nn \\
L^{-3} f_4&=\hat{\beta} \hat{C}_1 h_8(\hat{y})D\hat{\varphi}\wedge \dd \hat{y}\wedge \prim+\frac{h_8(\hat{y})}{4}D\psi\wedge \dd \hat{y}\wedge (\kappa_1 J_1+\kappa_2 J_2)\, , \nn \\
L^{-5}f_6&=- \frac{\kappa_1 \hat{b}_2+\kappa_2\hat{b}_1}{4} D\psi\wedge J_1\wedge J_2\wedge D\hat{\varphi}\, .
\end{align}
Note that for the solution to be well-defined we require
\be
\hat{b}_1>0\, ,\quad \hat{b}_2>0\, ,\quad C_1>0\, ,\quad \kappa_1\kappa_2\geq0\, .
\ee
There are therefore two choices one can make for the Riemann surface. Either one of the Riemann surfaces is a torus and the either is a torus or two-sphere or both are round two-spheres. We will see later in section \ref{sec:massiveGK} that this is in fact contained within a more general class of solution.

To bound the line interval parametrised by $\hat{y}$ we take the Romans mass to have jumps at positions $\hat{y}_i$, without loss of generality we can take $\hat{y}_0=0$ and take $h_8(0)=0$. This signifies the presence of an O8-plane which caps off the space. By allowing the Romans mass to have jumps at the $\hat{y}_i$ whilst keeping the function $h_8(\hat{y})$ continuous we may obtain a second root at $\hat{y}_{p+1}$ which bounds the space between a second O8-plane. The solution is thus compact and well-defined. Since we find that this solution is a specialisation of a more general solution we discuss later we will not present the quantisation of flux here and instead refer the reader to the later section \ref{sec:extremal}.

\subsubsection{Massless case}

Having considered the massive class of solution let us consider the massless solutions. We saw that in the massive case the solution was essentially unique, it turns out that this is not the case here and there is further branching. The $g_4$ Bianchi implies 
\be
h^{(1,1)}(y)=q_1\me^{-2 f_1(y)}=q_2\me^{-2 f_2(y)}\, ,
\ee
which has two solutions, either $h^{(1,1)}=0$ or $f_1(y)$ and $f_2(y)$ are proportional.

{\bf Case 1: $h^{(1,1)}=0$}\\
\noindent
Let us first consider the case where $h^{(1,1)}=0$. It follows that we must set
\be
\alpha=\beta=a_1=\kappa_2=0\, 
\ee
with the other parameters free. For the metric to have the correct signature we require that $\kappa_1=1$ and therefore we have a round $S^2$. In fact this combines with the R-symmetry direction to form a round $S^3$. Performing the rescalings
\begin{align}
m&\rightarrow L^{-1}\, ,\quad b_i\rightarrow \hat{b}_i\, ,\quad  a_2\rightarrow \frac{\hat{a}_2}{2L}\, ,\quad C_1\rightarrow L^{-2}\hat{C}_1\, ,\quad \delta\rightarrow \hat{\delta}\,, \quad \nonumber\\
 y&\rightarrow 2L\hat{y}\, ,\quad \varphi\rightarrow L^{3}\frac{\hat{\varphi}}{2}\, ,
\end{align}
the final solution takes the form
\begin{align}
\dd s^2&=L^2\sqrt{\hat{b}_1 \hat{\delta}}\bigg[\dd s^2(\text{AdS}_3)^{\text{unit}}+ \frac{1}{4}\dd s^2(S^3)
+\frac{(\hat{a}_2 \hat{y}+\hat{b}_2)}{4 \hat{b}_1}\Big(\hat{b}_1^2\hat{C}_1^2\dd\hat{\varphi}^2+\hat{b}_1\hat{C}_1^2 \hat{\delta}\dd y^2+\dd s^2(\mathbb{T}^2)\Big)\bigg]\, ,
\end{align}
with dilaton and magnetic fluxes
\begin{align}
\me^{4\Phi}&=\frac{\hat{b}_1^3 \hat{C}_1^4 \hat{\delta}^5(\hat{a}_2 \hat{y}+\hat{b}_2)^2}{256 }\, ,\quad 
&L^{-2}H&=\frac{a_2}{4}\dd\hat{\varphi}\wedge  \vol(\mathbb{T}^2)\, , \quad
f_2=0\, , \nonumber\\
L^{-3}f_4&=\frac{1}{2}\vol(S^3)\wedge \dd \hat{y}\, ,\quad
&L^{-5}f_6&=\frac{\hat{a}_2\hat{y}+\hat{b}_2}{8}\vol(S^3)\wedge \dd\varphi\wedge \vol(\mathbb{T}^2)\, .
\end{align}
This solution is in fact a special limit of a later solution and therefore we will not analyse it further. One can of course also quotient the $S^3$ with a subgroup of SU$(2)_R$ and preserve $\mathcal{N}=(2,0)$ supersymmetry.

{\bf Case 2: $h^{(1,1)}\neq 0$}

\noindent The second and final, class of massless solution allows for a non-trivial primitive two-form $H^{(1,1)}$. The final solution (after our favourite rescalings to make the metric coordinates and parameters dimensionless) is
\begin{align}
\dd s^2=&L^2\sqrt{\frac{\hat{b}_1 \hat{b}_2 \hat{\delta}}{\hat{b}_2\kappa_1+\hat{b}_1\kappa_2}}\bigg[\dd s^2(\text{AdS}_3)^{\text{unit}}+\frac{1}{4}D\psi^2+\hat{b}_1 \hat{b}_2 \hat{C}_1^2 D\varphi^2+\frac{ \hat{C}_1\hat{\delta}(\hat{b}_2 \kappa_1+\hat{b}_1\kappa_2)}{16}\dd y^2\nonumber\\
&+\frac{\hat{b}_2 \kappa_1+\hat{b}_1\kappa_2}{4 \hat{b}_2}\dd(\Sigma_1)^2+\frac{\hat{b}_2 \kappa_1+\hat{b}_1\kappa_2}{4 \hat{b}_1}\dd(\Sigma_2)^2\bigg]\, ,\\
D\varphi&\equiv\dd\varphi+\frac{\hat{\beta}}{\hat{b}_1\hat{b}_2}(\hat{b}_1 A_1-\hat{b}_2 A_2)\, ,\quad D\psi\equiv\dd\psi+\kappa_1 A_1+\kappa_2 A_2\, ,\nonumber
\end{align}
with dilaton and magnetic fluxes
\begin{align}
\me^{4\Phi}&=\frac{\hat{b}_1^3\hat{b}_2^3\hat{C}_1^4\hat{\delta}^5}{256 (\hat{b}_2\kappa_1+\hat{b}_1\kappa_2)}\, ,\quad L^{-2}H= \frac{\delta^2 \hat{q}_1 \hat{b}_2 \hat{C}_1^2}{16}\dd y\wedge (\hat{b}_1 J_1-\hat{b}_2 J_2)\, ,\quad L^{-1}f_2= \frac{\hat{q}_1}{\hat{b}_1}\Big(\hat{b}_2 J_2-\hat{b}_1 J_1)\, ,\nonumber\\
L^{-3}f_4&=-\hat{C}_1^2\beta \dd y \wedge D\varphi\wedge (\hat{b}_1 J_1-\hat{b}_2 J_2)+\frac{1}{4}D\psi \wedge \dd y \wedge (\kappa_1 J_1+\kappa_2 J_2)\, ,\\
L^{-5} f_6&= -\frac{\hat{b}_2 \kappa_1+\hat{b}_1\kappa_2}{4}D\psi\wedge D\varphi\wedge J_1\wedge J_2\, ,\nonumber
\end{align}
with 
\be
\hat{\beta}=\frac{\sqrt{4 \kappa_1\kappa_2 - \hat{q}_1^2 \hat{C}_1^2\hat{b}_2^2 \hat{\delta}^2}}{4\hat{C}_1}\, .
\ee
Note that setting $\hat{\beta}=\hat{q}_1=0$ leads to $\kappa_1\kappa_2=0$ and thus a solution in the previous section. 
Regularity imposes the inequalities
\be
\hat{b}_1\hat{b}_2>0\,,\quad 4\kappa_1\kappa_2-\hat{q}_1^2 \hat{C}_1^2\hat{b}_2^2 \hat{\delta}^2>0\, ,\quad \hat{\delta}(\hat{b}_2\kappa_1+\hat{b}_1\kappa_2)>0\, ,\quad (\hat{b}_2\kappa_1+\hat{b}_1\kappa_2)\hat{b}_1>0\, .
\ee
Clearly we require $\kappa_1\kappa_2>0$ unless $\hat{q}_1=\hat{\beta}=0$ in which case we have $\kappa_1\kappa_2=0$ as discussed in the previous section. Since $\hat{b}_1\hat{b}_2>0$ it follows that we must set $\kappa_1=\kappa_2=1$, otherwise we violate the last bound. The solution therefore contains two round two-spheres and turns out to be T-dual to a solution in the literature, namely the solution in section 3.1 of \cite{Donos:2008ug}.

%%%%%%%%%%%%%%%%%%%%%%%%%%%%%%%%%%
%%					Kahler ansatz
%%%%%%%%%%%%%%%%%%%%%%%%%%%%%%%%%%

\section{General K\"{a}hler base}\label{sec:GenKahler}

We will now assume that the conformally K\"ahler base is a non-trivial four-manifold, that is we take the $y$-dependence to come from an overall warp factor, so that
\be
\dd s^2\big(M_4(\vec{x},y)\big)=\me^{2 f(y)}\dd s^2(\mathcal{M}_4(\vec{x}))\, ,
\ee
with $\mathcal{M}_4$ K\"ahler. We may then decompose the SU$(2)$-structure forms as
\be
J=\me^{2 f(y)}J_{K}\, ,\qquad \Omega=\me^{2 f(y)}\Omega_K\, ,
\ee
where $J_K$ and $\Omega_K$ are the SU$(2)$-structure forms on $\mathcal{M}_4(\vec{x})$ and are independent of $y$. They satisfy
\be
\dd_4 J_K=0\, ,\qquad \dd_4 \Omega_K=\ii P_K\wedge \Omega_K\, ,
\ee
with $P_4$ the Ricci-form potential of the K\"ahler metric. We will allow the base to admit a closed primitive $(1,1)$-form which we denote by $\prim$.
Locally we may write it as
\be
\prim=\dd\Sigma\, ,
\ee
for some 1-form $\Sigma$ defined on $\mathcal{M}_4(\vec{x})$. Since the 2-form is primitive it implies that $\star_4 \prim=-\prim$ and consequently it is a harmonic 2-form. We will allow for all scalars to depend on both $y$ and the K\"ahler coordinates in the following. When a function does not depend on both sets we will explicitly give the coordinate dependence, but otherwise omit the arguments unless necessary. 

Our assumptions on the base and the torsion conditions for $J$ implies that we must set the primitive forms $T_2$ and $T_3$ to vanish. By construction equation \eqref{eq:Kahlercond} is satisfied whilst \eqref{eq:dyJ} imposes 
\be
\me^{ U}=g(\vec{x})\me^{ 2 f(y)+\tfrac{1}{2}(5 A -\Phi)}\, ,\label{eq:KahlerUsol}
\ee
with $g(\vec{x})$ an arbitrary non-zero function on the base.

From the torsion conditions for the holomorphic volume form we find,
\begin{align}
P&=P_K-\mathcal{A}_\varphi \sigma +\dd_4^c \log g(\vec{x})\, ,\\
\mathcal{A}_y &=-\mathcal{A}_\varphi \tau(\vec{x},y)\, .
\end{align}
We find that $\mathcal{A}_\varphi$ is constant and can therefore be removed by multiplying the holomorphic volume form by a $\varphi$-dependent phase. We may solve \eqref{eq:d4sigma} by taking
\be
T_1=\me^{U} t_1(\vec{x},y)\prim\, .
\ee
Integrability implies
\be
\dd_4 t_1(\vec{x},y)=0\, ,
\ee
and therefore
\be
\sigma= t_1(y)\Sigma\, .
\ee
From \eqref{eq:Riccifix} we find
\be
R_K+2\square_K \log g(\vec{x})=8 m^2g(\vec{x})^{-1} \me^{ \Phi-5A}\, ,
\ee
where we used \eqref{eq:KahlerUsol}. Note that the left-hand side is independent of $y$ and therefore we have that $\Phi-5A$ must be independent of $y$ too.

The Romans mass condition is equivalent to 
\be
\me^{2(\Phi-A)}f_0= 2 f'(y)+\partial_y(A-\Phi)\, .\label{eq:RomansKahler}
\ee
For the massless theory it is easy to see that this has solution
\be
\me^{\Phi-A}=c(\vec{x})\me^{2 f(y)}\, ,
\ee
with $c(\vec{x})$ a non-zero, but possibly constant, integration function. In the massive case for $f'(y)=0$ the general solution is 
\be
\me^{2(A-\Phi)}=2 f_0 y +b_1(\vec{x})\, ,
\ee
but more generally we can only solve this condition once we have fixed $f(y)$.

Let us proceed with the conditions from the Bianchi identities. 
We must fix the second primitive two-form $H^{(1,1)}$, which we take to be
\be
H^{(1,1)}=\me^{2f(y)+\tfrac{1}{2}(A+3\Phi)}h^{(1,1)}(\vec{x},y)\prim\, ,
\ee
then the Bianchi identity for the R--R 2-form implies the four conditions
\begin{align}
\dd_4^c\Big( g(\vec{x})\partial_y \me^{2 f(y)+A-\Phi}\Big)&=f_0 \dd_4^c\Big(g(\vec{x})\me^{2 f(y)+\Phi-A}\Big)\, ,\\
\dd_4\dd_4^c\Big(g(\vec{x})\me^{A-\Phi}\Big)+2 f_0 f'(y)J_K&=0\, ,\label{eq:g2cond2}\\
 \Big(\dd_4 \Big(\me^{\Phi-A}h^{(1,1)}(\vec{x},y)\Big)-t_1(y)\dd_4^c \big(g(\vec{x})\me^{A-\Phi}\big)\Big)\wedge \prim&=-f_0 t_1'(y) \Sigma\wedge J_K\, ,\label{eq:g2cond3}\\
t_1'(y)\Big(\dd_4 \big(g(\vec{x})\me^{A-\Phi}I\cdot\Sigma\big)+\dd_4^c \big(g(\vec{x})\me^{A-\Phi}\big)\wedge \Sigma\Big)&=
-\me^{\Phi-A}\Big(\partial_y h^{(1,1)}(\vec{x},y)+4 f'(y) h^{(1,1)}(\vec{x},y)\Big)\prim\, .\label{eq:g2cond4}
\end{align}
The first condition is satisfied immediately after using the condition for the Romans mass in \eqref{eq:RomansKahler}. The non-primitive part of the second condition implies the scalar condition
\be
\square_K\Big(g(\vec{x})\me^{A-\Phi}\Big)+4 f_0 f'(y)=0\,.
\ee
This is a necessary, but not sufficient condition since we must also enforce that the primitive part of the first term vanishes. For $f'(y)=0$ we see that we require a Laplacian Eigenfunction, but since the only Eigenfunctions on a K\"ahler manifold are constant it follows that the bracketed term is constant. We shall refrain from imposing any of these restrictions for the moment and proceed with the remaining Bianchi identities. 

For $H$ we find the two conditions
\begin{align}
0&=\Big(2 f''(y)+4 f'(y)^2\Big)J_K+\dd_4\dd_4^c \Big(g(\vec{x})\me^{\Phi-A}\Big)\, ,\\
0&=\Big(t_1''(y)+4 f'(y)t_1'(y)\Big) \Sigma\wedge J_K -\bigg[\dd_4 \Big(\me^{3 (\Phi-A)}h^{(1,1)}(\vec{x},y)\Big)+t_1(y)\dd_4^c \Big(g(\vec{x})\me^{\Phi-A}\Big)\bigg]\wedge \prim\, .
\end{align}
In the massive case these are implied by the Bianchi identity conditions for $g_2$, but in the massless case they are generically not. 

Finally let us consider the Bianchi identity for the R--R 4-form. There are three conditions, the first two are
\begin{align}
\Big[t_1(y)\dd_4\Big(g(\vec{x})\me^{A-\Phi}I\cdot\Sigma\Big)+t_1'(y)\Sigma\wedge \dd_4^c\Big(g(\vec{x})\me^{3 A}\Big)\Big]\wedge J&=0\, ,\\
g(\vec{x})\Big[ 2 f'(y) t_1'(y)\me^{A-\Phi}+\me^{-4 f(y)}\partial_y \big(\me^{4 f(y)+A-\Phi}t_1'(y)\big)\Big]I\cdot \Sigma \wedge J_K&=\label{eq:g4cond2}\\
\Big[ t_1(y)\dd_4 g(\vec{x})^2- g(\vec{x})&h^{(1,1)}(\vec{x},y)\dd_4^c\big(\me^{2(\Phi-A)}\big)\Big]\wedge\prim\, .\nonumber
\end{align}
whilst the final condition from the Bianchi identity gives the 4-form equation
\begin{align}
0&=\frac{\me^{-4 f(y)}}{g(\vec{x})^2}\partial_y\Big(\me^{-4 A}\big(2 f'(y)+\partial_y( 5 A -\Phi)\big)\Big) \frac{1}{2}J_K\wedge J_K+\Big[\me^{4 (\Phi-A)}\big(h^{(1,1)}(\vec{x},y)\big)^2+t_1(y)^2 g(\vec{x})^2\Big]\prim\wedge \prim\nonumber\\
&-\me^{-4 f(y)}\dd_4\dd_4^c \Big(g(\vec{x})^{-1}\me^{\Phi-5 A}\Big)\wedge J_K +2 t_1'(y)^2 g(\vec{x})\me^{A-\Phi}\Sigma\wedge I\cdot\Sigma\wedge J_K\nonumber\\
&+\frac{\me^{-4 f(y)}}{4 m^2}\Big(\dd_4 P_K\wedge \dd_4 P_K+2 \dd_4 P_K\wedge \dd_4\dd_4^c \log g(\vec{x})+\dd_4\dd_4^c \log g(\vec{x})\wedge \dd_4\dd_4^c \log g(\vec{x})\Big)\, .
\end{align} 
We may rewrite this as a scalar equation:
\begin{align}
&\square_K R_K-\frac{1}{2}R_K^2 +R^{mn}R_{mn}+2\square_K \square_K \log g(\vec{x})+2 R \square_K \log g(\vec{x})-4 R^{mn}\nabla_{m}\nabla_n\log g(\vec{x})\nonumber\\
&-2 (\square_K \log g(\vec{x}))^2 +2 \nabla^{m}\nabla^{n}\log g(\vec{x})\Big(\nabla_{m}\nabla_n\log g(\vec{x})-\tensor{I}{_{n}^{r}}\nabla_{r}\nabla_{s}\log g(\vec{x})\tensor{I}{^{s}_{m}}\Big)\label{eq:newmaster}\\
&=8 m^2 \me^{4 f(y)}\bigg[ 2 t_1'(y)^2 g(\vec{x})\me^{A-\Phi}|\Sigma|^2 +\frac{2\me^{-4 f(y)}}{g(\vec{x})^2}\partial_{y}\Big(\me^{-4 A}f'(y)\Big)-\Big[ \me^{4 (\Phi-A)}(h^{(1,1)}(\vec{x},y))^2+t_1(y)^2 g(\vec{x})^2\Big]|\prim|^2\bigg]\nonumber
\end{align}

Above we have presented results for a general K\"ahler base with a generic one-form $\Sigma$ and its closed primitive two-form field-strength $\prim$. We now want to solve the conditions explicitly whilst keeping the generality of our ansatz by not inserting any particular K\"ahler metric. A useful restriction is to consider four-dimensional toric metrics, this preempts our later discussion. For four-dimensional toric metrics the one-form $\Sigma$ is $\dd^c_4$-closed, in fact it can be written as $\dd^c_4 s$ for some function $s$. In addition we can introduce symplectic coordinates. This allows us to split the form equations depending on the number of legs along the torus coordinates of the toric action. For example $\Sigma$ has legs only along the angular coordinates and none along the non-angular coordinates. In addition, note that since $\prim$ is primitive it is anti-self-dual and therefore $\prim\wedge \prim\neq0$.

\subsection{Massive GK geometries}\label{sec:massiveGK}

In the previous section we have further reduced the conditions for a solution to exist with a warped K\"ahler metric. In this section we will solve these conditions explicitly, focussing first on the massive case. From \eqref{eq:g2cond3} we see that necessarily
\be
\dd_4\Big(\me^{\Phi-A}h^{(1,1)}(\vec{x},y)\Big)=0\, ,
\ee
whilst \eqref{eq:g2cond4} implies
\be
\partial_y h^{(1,1)}(\vec{x},y)+4 f'(y)h^{(1,1)}(\vec{x},y)=0\, , \qquad t_1'(y)\dd_4^c\big( g(\vec{x})\me^{A-\Phi}\big)\wedge \Sigma=0\, .
\ee
From \eqref{eq:g2cond3} we also have
\be
t_1(y)\dd_4^c \big(g(\vec{x})\me^{A-\Phi}\big)\wedge \prim =f_0 t_1'(y)\Sigma\wedge J_K\, ,
\ee
and combining with the previous condition we find that we should impose the two conditions
\be
t_1'(y)=0\, ,\quad \dd_4\big(g(\vec{x})\me^{A-\Phi}\big)=0\, .\label{eq:reducedt1'}
\ee
From \eqref{eq:g2cond2} it then follows that $f'(y)=0$ and therefore we have
\be
\me^{2(A-\Phi)}=2 f_0 y+c(\vec{x})\,.
\ee
It follows from \eqref{eq:reducedt1'} that $\dd_4 c(\vec{x})=\dd_4 g(\vec{x})=0$ and the only non-trivial remaining conditions fix $h^{(1,1)}(\vec{x},y)$ to be constant. It remains to solve the final scalar constraint in \eqref{eq:newmaster}. The condition reduces to
\begin{align}
\square_K R_K-\tfrac{1}{2}R_K^2+R^{mn}R_{mn}=-8 m^2 \me^{4 f}\Big[(2 f_0 y +b_1)^{-2}(h^{(1,1)})^2+ t_1^2 g^2\Big]|\prim|^2\, .
\end{align}
First note that the left-hand side is the same master equation components that govern GK geometries \cite{Gauntlett:2007ts}, whilst the right-hand side acts as a flux source term. Since the left-hand side is independent of $y$ it follows that we must set $h^{(1,1)}=0$. 

The final solution is
\begin{align}
\dd s^2&=\me^{2A}\bigg[\dd s^2(\text{AdS}_3)+\frac{1}{4m^2}(\dd\psi+P_{K})^2+(\dd\varphi+t_1 \Sigma)^2+\me^{\Phi-5A}\Big(\me^{\Phi-A}\dd y^2+\dd s^2(\mathcal{M}_4)\Big)\bigg]
\end{align}
where
\be
\me^{-4A}=\frac{R_K}{8 m^2}\sqrt{2 f_0 y+c}\, ,\qquad \me^{-2\Phi}=\frac{(2 f_0 y+c)^{5/4} \sqrt{R_K}}{2\sqrt{2} m}\, ,
\ee
where the metric on $\mathcal{M}_4$ is K\"ahler and satisfies the master equation 
\be
\square_K R_K-\tfrac{1}{2}R_K^2+R^{mn}R_{mn}=8 m^2 t_1^2 |\dd\Sigma|^2\, .\label{eq:master}
\ee
The solution is supported by the magnetic fluxes
\begin{align}
H&=0\, ,\quad
f_0\, ,\quad 
f_2=0\, , \nonumber\\
f_4&=t_1 D\varphi\wedge \dd y \wedge \prim+\frac{1}{4m^2}\Big((\dd\psi+P_K) \wedge \dd P_K-\frac{1}{2}\star_4 \dd R_K\Big)\wedge \dd y\, ,\\
f_6&=-\frac{1}{4 m^2}(\dd\psi+P_K)\wedge D\varphi\wedge J \wedge \dd P_K\, .\nonumber
\end{align}
We have presented a general class of solution above determined by solving the master equation, \eqref{eq:master}, for a four-dimensional K\"ahler base. Note that there is a large similarity between the geometry here and the so-called GK geometries \cite{Gauntlett:2007ts} that appear in AdS$_3$ solutions of Type IIB \cite{Kim:2005ez} and AdS$_2$ solutions of 11d supergravity \cite{Kim:2006qu}. One of the advances made in investigating these solutions is the construction of an extremal problem that determines the central charge (IIB)/free energy (11d) of the solution using just the topology of the manifold and without the need for an explicit metric, see \cite{Couzens:2018wnk}. Given the close connection we can also define an extremal problem for our setup, and the first in massive Type IIA.

\subsection{The extremal problem}\label{sec:extremal}

In the remainder of this section we will set $t_1=0$. To put the solution into a more amenable parametrisation for the extremal problem we first perform a few redefinitions. First define a new length scale $L=m^{-1}$ and redefine 
\begin{align}
\dd s^2(\text{AdS}_3)&=L^2 \dd s^{2}(\text{AdS}_3)^{\text{unit}}\, ,\quad \dd s^2(\mathcal{M}_4)=L^2\dd s^2(\mathcal{B}_4)\, ,\quad \varphi= L\hat{\varphi}\, ,\quad f_0=\frac{\hat{f}_0}{L}\, ,\nonumber\\
 y&=L\frac{-c+h_8(\hat{y})^2}{2\hat{f}_0}\, ,\quad  h_8(\hat{y})=\hat{f_0}\hat{y}+\hat{c}\, .
\end{align}
It is also useful to define
\be
\me^{B}=\frac{R_B}{8}\, ,\quad \eta=\frac{1}{2}(\dd\psi+P)\, ,
\ee
with $R_B$ the (dimensionless) Ricci scalar of the base metric $\mathcal{B}_4$.
With these redefinitions the metric becomes
\be
L^{-2}\dd s^2=\frac{\me^{-B/2}}{\sqrt{h_8(\hat{y})}}\bigg[\dd s^2(\text{AdS}_3)^{\text{unit}}+\eta^2+\dd \hat{\varphi}^2+\me^{B}\Big(\dd s^2(\mathcal{B}_4)+h_8(\hat{y})\dd\hat{y}^2\Big)\bigg]\, .
\ee
We have dropped the subscript ``$K$" on the 1-form $P$, and all forms will be the defined on $\mathcal{B}_4$ henceforth. Recall that $\dd P=\rho$ with $\rho$ the Ricci-form of $\mathcal{B}_4$. The metric is then in a form similar to GK geometries, and satisfies the same master equation, this time in four dimensions. The 1-form $\eta$ is the 1-form dual to R-symmetry Killing vector $\partial_{\psi}\equiv\xi$ and in keeping with the notation in \cite{Couzens:2018wnk} we call this the R-symmetry vector. We define $Y_5$ to be the manifold consisting of the U$(1)$ R-symmetry direction fibered over the base $\mathcal{B}_4$. This is the five-dimensional version of a GK geometry. 

We can identify the linear function $h_8(\hat{y})$ as the warp factor of D8-branes which are arrayed at fixed points along the $\hat{y}$ line interval and wrap the remaining directions. After the redefinitions the non-trivial dilaton and magnetic fluxes take the form
\begin{align}
\me^{-2\Phi}&=h_8(\hat{y})^{5/2}\me^{B/2}\, ,\\
L \, f_0&=\partial_{\hat{y}}h_{8}(\hat{y})\, ,\\
L^{-3} f_4&= \frac{h_8(\hat{y})}{2}\Big(\eta\wedge \rho-\frac{1}{4}\star_4 \dd R\Big)\wedge \dd\hat{y}\, ,\\
L^{-5}f_6&=-\frac{1}{2}\eta\wedge \rho\wedge J\wedge \dd\hat{\varphi}\, .
\end{align}

For the solution to be well-defined it remains to fix the period of $\hat{y}$ and quantise the fluxes correctly. 
Supersymmetry and the equations of motion impose that $h_8(\hat{y})$ is a linear function with first derivative the Romans mass. The function $h_8(\hat{y})$ must be continuous but need only be piecewise smooth, in particular we may allow for jumps in the Romans mass in different patches of the line interval, let there be $p$ such jumps. Note that we require $p\geq 1$ for the space to close, if there is no jump the space is non-compact. We can then bound the line interval between two zeroes of $h_8(\hat{y})$. Without loss of generality we may take the smaller root to be at $0$ and the second to be at some strictly positive root $\hat{y}_{p+1}$ which caps the space. At the two end-points the degeneration of the metric shows that the space is capped off by O8-planes. 

The most general $h_8$ one may construct is\footnote{We include an overall factor depending on the string length for later.}
\be
h_8(\hat{y})=\frac{2\pi \ls}{L}\begin{cases}
\hat{f}_0^{(0)}\hat{y}& 0\leq \hat{y}\leq \hat{y}_1\, ,\\
\quad \vdots\\
\hat{f}_0^{(i)}(\hat{y}-\hat{y}_i)+ c^{(i)}&\hat{y}_i\leq \hat{y}\leq\hat{y}_{i+1}\, ,\\
\quad\vdots\\
\hat{f}_0^{(p)}(\hat{y}-\hat{y}_{p})+c^{(p)}&\hat{y}_p\leq \hat{y}\leq\hat{y}_{p+1}\, ,
\end{cases}
\ee
subject to the continuity conditions
\begin{align}
\hat{f}_0^{(i)}(\hat{y}_{i+1}-\hat{y}_{i})+c^{(i)}&=c^{(i+1)}\, ,\\
\hat{f}_0^{(p)}(\hat{y}_{p+1}-\hat{y}_{p})+c^{(p)}&=0\, .
\end{align}
Note that this defines the $c^{(i)}$ iteratively as
\be
c^{(i+1)}= \sum_{k=0}^{i}\hat{f}_0^{(k)}(\hat{y}_{k+1}-\hat{y}_{k})\, ,
\ee
and a constraint on $\hat{y}_{p+1}$,
\be
0=\sum_{k=0}^{p}\hat{f}_{0}^{(k)}(\hat{y}_{k+1}-\hat{y}_{k})\, .
\ee
In addition we require that $h_8(\hat{y})$ defines a convex curve, this implies 
\be
\hat{f}_0^{(i)}>\hat{f}_0^{(i+1)}\, ,\qquad \forall i
\ee
and guarantees that only D8-branes appear in the bulk as opposed to O8-planes. There are $2 p+1$ free parameters, the $p+1$ constants $\hat{f}_0^{(i)}$ and the $p$ locations of a jump $y_{i}$, $1\leq i\leq p$. The final end-point is fixed by the choice of this data. In addition as we will see soon this data is further constrained by flux quantisation. 

We now want to rephrase the problem of computing the central charge and performing flux quantisation as an extremal problem following \cite{Couzens:2018wnk}. Note that the geometry $Y_5$ considered here is precisely the $n=2$ version of the theory considered in \cite{Gauntlett:2007ts,Couzens:2018wnk}. Our solution is determined by a $2p+1$-dimensional charge vector containing the D8-brane information, the $\hat{f}_{0}^{(i)}$ and $\hat{y}_{i}$ parameters, and a base $Y_5$ which is of GK type \cite{Gauntlett:2007ts} for $n=2$. As such the extremal problem will make use of existing results in the literature, in particular in \cite{Couzens:2018wnk} and the followups \cite{Gauntlett:2019pqg,Kim:2019umc,Hosseini:2019ddy,Gauntlett:2019roi,Zaffaroni:2019dhb,Hosseini:2019use,Gauntlett:2018dpc,vanBeest:2020vlv}. For clarity we will review these results repurposed to our problem.  

We first fix the complex cone $C(Y_5)$ and endow it with a nowhere-zero closed holomorphic three-form and holomorphic U$(1)^s$ action. We pick a basis of the U$(1)^s$ action where the holomorphic volume form has charge 2 under the first basis vector and is uncharged under the remaining $s-1$ vectors. The R-symmetry vector may then be written as
\be
\xi=\sum_{I=1}^{s}b_I\partial_{\psi_I}\, .
\ee
The vector $\vec{b}\equiv(b_1,b_2,b_3)$ parametrises the choice of R-symmetry vector, and is subject to $b_1=2$ which should be imposed at the end. 
We may define the 5d supersymmetric action 
\be
S_{\text{SUSY}}[\xi,J]=\int_{Y_5}\eta\wedge \rho\wedge J\, ,
\ee
which is a functional of the choice of R-symmetry and K\"ahler metric. Note that it depends only on the cohomology class and not the explicit representative. The master equation \eqref{eq:master} (with $t_1=0$) may be integrated to obtain
\be
0=\int_{Y_5}\eta\wedge \rho \wedge \rho\,,\label{eq:intmaster}
\ee
which is a necessary condition for the equations of motion to be solved. With the above constraint and the assumption that the cohomology condition
\be
H^2(Y_5,\mathbb{R})\simeq H_{\mathcal{B}_4}^2(\mathcal{F}_{\xi})/[\rho]\, ,
\ee
holds true, where $\mathcal{F}_{\xi}$ is the transverse foliation of the R-symmetry vector, we may define a consistent quantisation of the fluxes.

The quantisation condition for the magnetic fluxes is
\be
2\pi \ls f_0=n\in\mathbb{Z}\, ,\quad \frac{1}{(2\pi \ls)^{3}}\int_{\sigma_{a,i}}f_4=M_{a,i}\in \mathbb{Z}\, ,\quad \frac{1}{(2\pi \ls)^{5}}\int_{\Sigma_A}f_6=N_A\in \mathbb{Z}\, ,
\ee
with $\sigma_{a,i}$ all four-cycles and $\Sigma_A$ all six-cycles in the geometry. The quantisation condition for the Romans mass implies 
\be
\hat{f}_0^{(i)}\in \mathbb{Z}\, ,
\ee
justifying our choice of normalisation earlier.
For the magnetic 6-form flux there is a single relevant six-cycle consisting of $Y_5$ and the U$(1)$ direction parametrised by $\hat{\varphi}$. Let $\hat{\varphi}$ have period $2\pi \ls l_{\varphi}/L$, then the quantisation condition reads
\be
\frac{L^4 l_{\varphi}}{2(2\pi \ls)^4}\int_{Y_5}\eta \wedge \rho\wedge J=N\in \mathbb{Z}\, ,
\ee
where we have dropped a total derivative term. This should be understood as the number of D2-branes in the geometry probed by the cone over $Y_5$ and smeared along the circle. We will turn to evaluating this integral using the results in \cite{Couzens:2018wnk} shortly. The final flux quantisation condition we must consider is the quantisation of the magnetic 4-form flux. The relevant four-cycles in the geometry consist of the union of line-segments with three-cycles in $Y_5$:
\be
\sigma_{a,i}=[\hat{y}_i,\hat{y}_{i+1}]\times\hat{\sigma}_{a}\, ,
\ee
with $\hat{\sigma}_{a}$ giving a basis of three-cycles in $Y_5$. In total there are $(p+1)\times b_3(Y_5)$ such cycles to consider. The quantisation condition becomes
\be
\bigg[\frac{L^2}{2(2\pi \ls)^2}\int_{\hat{\sigma}_a}\eta\wedge \rho\bigg]\times \bigg[\frac{L}{2\pi\ls}\int_{\hat{y}_i}^{\hat{y}_{i+1}}h_8(\hat{y})\dd\hat{y}\bigg]=M_{a,i}\in\mathbb{Z}\, ,
\ee
where we have split the terms suggestively.
The first integral may again be computed using the toric formulae in \cite{Couzens:2018wnk} as we will explain shortly. The second may be integrated to give:
\begin{align}
\frac{L}{(2\pi \ls)}\int_{\hat{y}_i}^{\hat{y}_{i+1}}h_8(\hat{y})\dd\hat{y}&=\frac{1}{2}(\hat{y}_{i+1}-\hat{y}_{i})(c^{(i+1)}+c^{(i)})\nonumber\\
&=\frac{1}{2}(\hat{y}_{i+1}-\hat{y}_{i})\bigg[\hat{f}_{0}^{(i)}(\hat{y}_{i+1}-\hat{y}_{i})+2\sum_{k=0}^{i-1}\hat{f}_{0}^{(k)}(\hat{y}_{k+1}-\hat{y}_{k})\bigg]
\end{align}
To satisfy this quantisation condition let us define
\be
\frac{L^2}{2(2\pi \ls)^2}\int_{\hat{\sigma}_a}\eta\wedge \rho=M_a\, ,
\ee
and
\be
\frac{L}{2\pi\ls}\int_{\hat{y}_i}^{\hat{y}_{i+1}}h_8(\hat{y})\dd\hat{y}=n_i\, ,
\ee
so that 
\be
M_{a,i}=M_a n_i\in \mathbb{Z}\, .
\ee
The simplest possibility to satisfy the condition is to take both $M_a$ and $n_i$ integer, however we may take the more general choice of fixing $M_a\in\mathbb{Z}$ and 
\be 
n_i\in \mathbb{Z}/\text{gcd}(M_a)\, .
\ee
We see that imposing flux quantisation splits into a part dependent on the geometry of $Y_5$ and a second part dependent only on the D8-brane parameters. 
We then need to use the results in \cite{Couzens:2018wnk,Gauntlett:2018dpc} to evaluate the following integrals for flux quantisation
\begin{align}
\frac{L^4 l_{\varphi}}{2(2\pi \ls)^4}\int_{Y_5}\eta \wedge \rho\wedge J&=N\in \mathbb{Z}\, ,\\
\frac{L^2}{2(2\pi \ls)^2}\int_{\hat{\sigma}_a}\eta\wedge \rho&=M_a\in \mathbb{Z}\, .
\end{align}

Next observe that the central charge of the solution can be obtained by using the Brown--Henneaux formula \cite{Brown:1986nw} giving
\begin{align}
c&=\frac{48 \pi^2 L^8}{(2\pi\ls)^8}\int_{M_7}h_8(\hat{y})\me^{B}\frac{1}{2}\eta\wedge J^2\wedge \dd\hat{\varphi}\wedge \dd\hat{y}\nonumber\\
&=\frac{12 \pi^2 L^8}{(2\pi\ls)^8}\int_{M_7}h_8(\hat{y})\eta\wedge J\wedge \rho \wedge \dd\hat{\varphi}\wedge \dd\hat{y}\, ,
\end{align}
where in the last line we have used the properties of the Ricci-form. In terms of the supersymmetric 5d action this is
\be
c=\frac{12 \pi^2 L^7 l_{\varphi}}{(2\pi\ls)^7} S_{\text{SUSY}}\int h_8(\hat{y})\dd\hat{y}\, .
\ee
Strictly this is off-shell expression for the central charge as we have not satisfied the equations of motion yet. 

Once the constraint equation \eqref{eq:intmaster} and flux quantisation have been imposed we can then extremise the supersymmetric action $S_{\text{SUSY}}$ over the choice of R-symmetry vector and K\"ahler parameters. The central charge is then
\be
c=\frac{12 \pi^2 L^7 l_{\varphi}}{(2\pi\ls)^7} S_{\text{SUSY}}\Big|_{\text{on-shell}}\int h_8(\hat{y})\dd\hat{y}\, .
\ee
Using the flux quantisation we may rewrite this as
\be
c=\frac{48\pi^2 N M_1}{\int_{\hat{\sigma}_1}\eta\wedge \rho}\sum_{i=0}^{p} n_i\, ,
\ee
where we picked a particular flux number $M_1$. Note the similarity with the expression in \cite{Couzens:2018wnk}. The presence of the D8-branes lead to a deformation of the central charge by the $n_i$ dependent piece. One must still impose that the flux parameters $M_a$ are related as 
\be
\frac{M_1}{M_{a\neq1}}=\frac{\int_{\hat{\sigma}_1}\eta\wedge \rho}{\int_{\hat{\sigma}_a}\eta\wedge \rho}\, .
\ee

We now want to evaluate the final integrals for flux quantisation. We may directly use the expressions in \cite{Couzens:2018wnk} if we assume that $Y_5$ is toric or the more elegant expressions using the master volume in \cite{Gauntlett:2018dpc}. This was later extended to toric manifolds fibered over a K\"ahler base in \cite{Gauntlett:2019pqg} and one could in principle use their results for an $S^3$ fibered over a general Riemann surface or an $S^1$ over a 4d K\"ahler base.\footnote{Strictly these should be free of orbifold singularities.} Rather than presenting these more complicated cases we will present the simpler case when $Y_5$ is toric for completeness. 

If $Y_5$ is toric it means we have a holomorphic U$(1)^s$ action. This defines a set of vectors $v_a$, $a=1,...,d$ which are inward pointing normals to the facets of the polyhedral cone and define the geometry. We refer the reader to \cite{abreu2000kahler} for a more detailed exposition of toric geometry. One can define the master volume
\be
\mathcal{V}\equiv \int_{Y_5}\frac{1}{2}\eta\wedge J\wedge J\, .
\ee
The K\"ahler form may be expanded in a basis $C_a$ of basic representatives of $H^{2}_{\mathcal{B}_4}(\mathcal{F}_\xi)$, see \cite{abreu2000kahler,Gauntlett:2018dpc}, where the two-forms $C_a$ are Poincar\'e dual to the restriction of toric divisors of the cone $C(Y_5)$, as
\be
[J]=-2\pi \sum_{a=1}^{d}\lambda_a C_a\, .
\ee
Only $d-3$ of the $C_a$ are independent \cite{abreu2000kahler}, and thus only $d-3$ of the K\"ahler parameters $\lambda_a$ will appear in the expressions. The Ricci form can be expanded similarly as
\be
\rho=2\pi \sum_{a=1}^{d}C_a\, .
\ee
The master volume of $Y_5$ can then be determined in terms of the toric data as
\be
\mathcal{V}(\vec{b},\lambda_a,\vec{v}_a)=\frac{(2\pi)^3}{2}\sum_{a=1}^{d}\lambda_a \frac{\lambda_{a-1}(\vec{v}_a,\vec{v}_{a+1},\vec{b})-\lambda_a (\vec{v}_{a-1},\vec{v}_{a+1},\vec{b})+\lambda_{a+1}(\vec{v}_{a-1},\vec{v}_{a},\vec{b})}{(\vec{v}_{a-1},\vec{v}_{a},\vec{b})(\vec{v}_a,\vec{v}_{a+1},\vec{b})}\, .
\ee
It then follows that the integrals we needed for flux quantisation and the constraint equation may be determined in terms of the master volume as
\begin{align}
\int_{Y_5}\eta\wedge \rho \wedge\rho&=\sum_{a,b=1}^{d}\frac{\partial^2 \mathcal{V}}{\partial \lambda_{a}\partial \lambda_{b}}\, ,\\
\int_{Y_5}\eta\wedge \rho\wedge J &=-\sum_{a=1}^{d} \frac{\partial \mathcal{V}}{\partial \lambda_{a}}\, ,\\
\int_{\hat{\sigma}_{a}}\eta\wedge \rho&=\frac{1}{2\pi}\sum_{b=1}^{d}\frac{\partial^2 \mathcal{V}}{\partial \lambda_{a}\partial \lambda_{b}}\, .
\end{align}

With the above expressions we may perform the quantisation of the fluxes, extremise the action and obtain the central charge. First we determine the constraint. Note that since the master volume is quadratic in $\lambda_a$ the constraint 
\be
\sum_{a,b=1}^{d}\frac{\partial^2 \mathcal{V}}{\partial \lambda_{a}\partial \lambda_{b}}=0\, ,
\ee
is independent of $\lambda$ and must be solved for $b_3$ (or $b_2$). We can solve for the flux parameter $N$ in terms of one of the $\lambda_a$,
\be
N=-\frac{2(2\pi \ls)^4}{L^4 l_{\varphi}}\sum_{a=1}^{d} \frac{\partial \mathcal{V}}{\partial \lambda_{a}}
\ee
Next the fluxes $M_a$ are given by
\be
M_a=\frac{2(2\pi \ls)^2}{2\pi L^2}\sum_{b=1}^{d}\frac{\partial^2 \mathcal{V}}{\partial \lambda_{a}\partial \lambda_{b}}\, ,
\ee
and are independent of the $\lambda_a$. We must now impose that 
\be
M_1 \sum_{b=1}^{d}\frac{\partial^2 \mathcal{V}}{\partial \lambda_{a}\partial \lambda_{b}}=M_a \sum_{b=1}^{d}\frac{\partial^2 \mathcal{V}}{\partial \lambda_{1}\partial \lambda_{b}}\, ,\label{eq:fluxequate}
\ee
for all $a$. This imposes $b_3(Y_5)-1$ constraints for a single free variable. If $b_3(Y_5)=1$ we retain the free variable $b_2$ which is fixed by extremising the trial central charge
\be
c=\frac{96\pi^3 N M_1}{\sum_{b=1}^{d}\frac{\partial^2 \mathcal{V}}{\partial \lambda_{1}\partial \lambda_{b}}}\sum_{i=0}^{p} n_i\, ,
\ee
over the remaining parameter $b_2$.\footnote{This is independent of the $\lambda$'s again.} If $b_3(Y_5)=2$ then we may use \eqref{eq:fluxequate} to fix $b_2$ in terms of the flux parameters, and then there is nothing left to extremise. For $b_3(Y_5)>2$ not only is there nothing to extremise and $b_2$ is fixed in terms of the fluxes once again but the fluxes $M_a$ also satisfy a non-trivial constraint. This agrees with the analysis in \cite{Couzens:2018wnk} of 7d GK geometries with a $T^2$ factor. 

{\bf Brane construction}

The gravity solution suggests the brane realisation given in table \ref{Table:Dbrane}. This consists of a stack of $N$ D2-branes probed by the cone over $Y_5$, with flavour D4 and D8-branes located at the distinguished points along the line interval. One should think of this brane construction as giving rise to a quiver field theory consisting of subquivers joined together and flavoured by D8-branes. Each subquiver is given by the field theory living on a stack of $N$ D2-branes probed by $C(Y_5)$, smeared along a circle and wrapping a finite length line interval. It would be interesting to construct such a dual field theory in the future. 

\begin{table}
\begin{center}
\begin{tabular}{|c||c|c|c|c|c|c|c|c|c|c|}
\hline
Brane&\multicolumn{2}{c|}{$\mathbb{R}^{1,1}$}&\multicolumn{6}{c|}{$C(Y_5)$}&$U(1)$& $I$ \\
\hline
 &  \multicolumn{2}{c|}{}&\multicolumn{4}{c|}{}&  \multicolumn{2}{c|}{$C$}& &\\
\hline
\hline
D2&$\times$&$\times$&$\bullet$&$\bullet$&$\bullet$&$\bullet$&$\bullet$&$\bullet$&$-$&$\times$\\
\hline
D4& $\times$&$\times$&$-$&$-$&$-$&$-$&$\times$&$\times$&$\times$&$\bullet$\\
\hline
D8&$\times$&$\times$&$\times$&$\times$&$\times$&$\times$&$\times$&$\times$&$\times$&$\bullet$\\
\hline
\end{tabular}
\caption{The brane configuration arising from our setup. The D2-branes are located at the tip of the cone over $Y_5$, smeared over the circle and lie along the line interval. The D4-branes wrap a two-cycle in $Y_5$, denoted by $C$ in the table, along with the circle and are located at the special points of the line interval. Finally the D8-branes are located at the distinguished points of the line interval.}
\label{Table:Dbrane}
\end{center}
\end{table}

Another interesting point to highlight is the connection of these geometries to the F-theory ones discussed in \cite{Couzens:2017nnr,Passias:2019rga,Couzens:2019iog} and for which an extremal problem was derived in \cite{vanBeest:2020vlv} by making use of M/F-duality. Dualising along the U$(1)$ to Type IIB the D8-branes become 7-branes and the Romans mass becomes a non-trivial axion. The metric is then of the class discussed in section 4.2.2 of \cite{Couzens:2017nnr} and should give a local description of the base of an elliptically fibered K3 surface or equivalently locally is the hyper-K\"ahler manifold
\be
\dd s^2(\text{HK})=h_8(\hat{y})\big(\dd y^2+\dd x_1^2+\dd x_2^2\big)+\frac{1}{h_8(\hat{y})}\big(\dd \hat{\varphi}+h_8'(\hat{y}) x_1\dd x_2\big)^2\, .
\ee
It would be interesting to map the extremal problem here into the one considered in \cite{vanBeest:2020vlv}.

\section{Geometric conditions for arbitrary extended chiral supersymmetry}\label{sec:extendedsusy}
In this section we give necessary and sufficient conditions for AdS$_3$ solutions of Type II supergravity to preserve arbitrary extended chiral supersymmetry --- i.e.\ ${\cal N}=(n,0)$ for $2<n\leq 8$  (The case of ${\cal N}=(8,0)$ is maximal \cite{Haupt:2018gap}). As we shall see, these conditions are actually implied by the ${\cal N}=(2,0)$ conditions of section \ref{sec:SUSY}.\\
~\\
A solution preserving ${\cal N}=(n,0)$ supersymmetry must support two $n$-tuplets of Majorana spinors on M$_7$, $\chi^I_{1,2}$ for $I=1,2,...,n$. In terms of these one can define $\frac{n}{2}(n-1)$ independent ${\cal N}=2$ sub-sectors, the same number of independent components as a dim($n$) anti-symmetric matrix. Each of these sub-sectors must obey the conditions of section \ref{sec:SUSY} for potentially differing $(\xi,\tilde \xi, c,\Psi_{\pm})$. However, by exploiting constant GL($n$,$\mathbb{R}$) transformations of $\chi^I_{1,2}$ one can take them to obey
\beq
\chi^{I\dag}_1\chi^{J}_1+\chi^{I\dag}_2\chi^{J}_2= 2e^{A}\delta^{IJ}\, ,~~~~\chi^{I\dag}_1\chi^{J}_1-\chi^{I\dag}_2\chi^{J}_2= c e^{-A}\delta^{IJ}\, ,
\eeq
without loss of generality --- so that  they share the same $c$. In terms of these one can define the real 1-form valued $n\times n$ anti-symmetric matrices
\begin{align}
\xi^{IJ}&= -i(\chi^I_1\gamma_a\chi^J_1 \mp  \chi^I_2\gamma_a\chi^J_2)\e^a,~~~\tilde\xi^{IJ}= -i(\chi^I_1\gamma_a\chi^J_1 \pm  \chi^I_2\gamma_a\chi^J_2)\e^a\, ,
\end{align}
where the vectors dual to the components of $\xi^{IJ}$ are all Killing vectors with respect to the entire solution under which $\chi^I_{1,2}$ are charged. The necessary and sufficient conditions for ${\cal N}=(n,0)$ supersymmetry can then be expressed covariantly as
\begin{subequations}
\begin{align}
&d\tilde\xi^{IJ}= \iota_{\xi^{IJ}}H~~~~e^{3A}h_0=-m c\, ,\label{eq:genbps0}\\[2mm]
&d_{H}(e^{A-\Phi}\Psi^{(IJ)}_{\mp})=\mp \frac{c}{16}\delta^{IJ}f_{\pm}\, ,\label{eq:genbps1}\\[2mm]
&d_{H}(e^{2A-\Phi}\Psi^{(IJ)}_{\pm})\mp 2m e^{A-\Phi}\Psi^{(IJ)}_{\mp}= \frac{1}{8}e^{3A}\star_7\lambda f_{\pm} \delta^{IJ}\, ,\label{eq:genbps2}\\[2mm]
&d_{H}(e^{-\Phi}\Psi^{[IJ]}_{\pm})=\frac{1}{16}\big(\tilde{\xi}^{IJ}\wedge+\iota_{\xi^{IJ}}\big) f_{\pm}\, ,\label{eq:genbps3}\\[2mm]
&d_{H}(e^{3A-\Phi}\Psi^{[IJ]}_{\mp})=\mp e^{3A-\Phi}h_0\Psi^{[IJ]}_{\pm}\pm \frac{1}{16}\big(\tilde{\xi}^{IJ}\wedge +\iota_{\xi^{IJ}}\big)e^{3A}\star_7 \lambda f_{\pm}\label{eq:genbps4}\, ,\\[2mm]
&(\Psi^{(IJ)}_{\mp},f_{\mp})_7=\mp\frac{1}{2}\delta^{IJ}\left(m+\frac{1}{4}e^{-A}ch_0\right)e^{-\Phi}\text{vol}(\text{M}_7)\, ,\label{eq:genbps5}
\end{align}
\end{subequations}
which we should stress contain many redundant expressions. What considering the $\frac{n}{2}(n-1)$ independent ${\cal N}=(2,0)$ sub-sectors does not tell us however is the following
\begin{enumerate}
\item How many of the $\frac{n}{2}(n-1)$ Killing vectors dual to $\xi^{IJ}$ are independent.
\item How $\chi_{1,2}^I$ transform under every component of $\xi^{IJ}$.
\end{enumerate}
To be clear we do know how the Killing vector associated to each ${\cal N}=(2,0)$ sub-sector acts on the spinors that make up that sector --- what we don't know is how they act on the remaining spinors of the $n$-tuplet.
If we were considering for instance AdS$_4$ solutions with extended supersymmetry, which have superconformal group  OSp($n|4$) and spinors transforming in the $\textbf{n}$ of the SO($n$)  R-symmetry, it would be clear that $\xi^{IJ}$ should contain all $\frac{n}{2}(n-1)$ independent  SO(n) Killing vectors (and possibly some additional flavour isometries). The structure of the chiral superconformal algebras for AdS$_3$ with extended supersymmetry is however more rich than the higher dimensional cases \cite{Beck:2017wpm}. There exists the analogous possibility of  OSp($n|2$) with spinors in the fundamental of the SO($n$) R-symmetry, however there are several other options. Several of these have R-symmetry groups of dimension less than $\frac{n}{2}(n-1)$, for small ${\cal N}=(4,0)$ for instance it is SU(2) . The algebras that are consistent with  AdS$_3$ solutions (the simple Lie super-algebras) can be classified in terms of the Lie algebra of their R-symmetry $\mathfrak{g}$ and a corresponding representation $\rho_{\mathfrak{g}}$ \cite{Fradkin:1992bz}, which $\chi^{I}_{1,2}$ should transform in under $\mathfrak{g}$. There should thus exist a real basis of $\mathfrak{g}$, $T_{\mathfrak{g}}^a$ for $a=1,...,\text{dim}(\mathfrak{g})$,  in the representation  $\rho_{\mathfrak{g}}$ such that
\beq
{\cal L}_{K_{\mathfrak{g}}^a}\chi^I_{1,2}=  (T_{\mathfrak{g}}^a)^{IJ}\chi_{1,2}^J\, ,
\eeq
where $K_{\mathfrak{g}}^a$  are the Killing vectors of $\mathfrak{g}$. This leads us to make  the mild conjecture that the different possibilities for super-conformal algebras can be distinguished by decomposing 
\beq\label{eq;conjecture}
\xi^{IJ}=  -8m c^a K_{\mathfrak{g}}^a (T_{\mathfrak{g}}^a)^{IJ}+ \xi_0^{IJ}\, ,~~~~{\cal L}_{\xi_0^{IJ}}\chi^L_{1,2}=0\, ,
\eeq
where $\xi_0^{IJ}$ are some additional flavour (or uncharged) isometries in M$_7$ that we cannot exclude the possibility of. Here $c^a$ are a set of constants one needs to keep arbitrary for consistency with large ${\cal N}=(4,0)$ which depends on a continuous parameter --- we expect $c_a=1$ in all other cases. Note we are assuming conventions where
\beq
[T_{\mathfrak{g}}^a,T_{\mathfrak{g}}^b]=-f^{abc}T_{\mathfrak{g}}^c\, ,~~~~ \text{Tr}(T_{\mathfrak{g}}^aT_{\mathfrak{g}}^b)=-\frac{n}{4}\delta^{ab}\, ,~~~~[K_{\mathfrak{g}}^a,K_{\mathfrak{g}}^b]=f^{abc}K_{\mathfrak{g}}^c\, ,
\eeq
which is the reason for the $-8m$ in the first expression in \eqref{eq;conjecture}. We have explicitly checked this proposal for the classes of small ${\cal N}=(4,0)$ solutions in \cite{Lozano:2019emq} and \cite{Macpherson:2022sbs}, the large ${\cal N}=(4,0)$ solutions in \cite{Macpherson:2018mif} and the ${\cal N}=(3,0)$ solutions in \cite{Legramandi:2019xqd}. In these cases one finds $\xi_0^{IJ}=0$, though we are aware of some examples for which this is not the case (see the discussion about a priori isometries in \cite{Macpherson:2022sbs}). 

\section*{Acknowledgments}
We would like to thank Salomon Zacarias. CC would like to thank Hyojoong Kim and Nakwoo Kim for discussions. CC is supported by the National Research Foundation of Korea (NRF) grant 2019R1A2C2004880.
NM  is supported by AEI-Spain (under project PID2020-114157GB-I00 and Unidad de Excelencia Mar\'\i a de Maetzu MDM-2016-0692), by Xunta de Galicia-Conseller\'\i a de Educaci\'on (Centro singular de investigaci\'on de Galicia accreditation 2019-2022, and project ED431C-2021/14), and by the European Union FEDER. AP is supported by the Hellenic Foundation for Research and Innovation (H.F.R.I.) under the ``First Call for H.F.R.I.
Research Projects to support Faculty members and Researchers and
the procurement of high-cost research equipment grant'' (MIS 1857, Project Number: 16519).

\appendix

\section{AdS$_3$ spinors and bi-linears}\label{sec:AdS3app}
In this appendix we will give some details on the spinors and bi-linears of AdS$_3$, which supplements the following appendix.\\
~~\\
AdS$_3$ is a maximally symmetric space with global SO(2,2) $=$ SL(2)$_+\times$SL(2)$_-$ symmetry and Ricci tensor ${\rm Ricci}(\text{AdS}_3)=-2m^2 g(\text{AdS}_3)$. It comes equipped with Killing spinors charged under SL(2)$_{\pm}$ defined through the Killing spinor equation
\beq
\nabla_{\mu}\zeta_{\pm}= \pm \frac{m}{2}\gamma^{(3)}_{\mu}\zeta_{\pm}\, ,
\eeq
where in this work we will be interested in solution preserving SL(2)$_+$ specifically. A particular parameterisation of AdS$_3$ is given by the vielbein
\beq\label{eq:vielAdS3}
\e^0= e^{mr}dt\, ,~~~~~\e^1= e^{mr} dx\, ,~~~~\e^2= dr\, ,
\eeq
and in terms of this, one can show that $\zeta_+$ decomposes to two independent components as
\beq
\zeta_+ =  c_1 \zeta^P+ c_2 \zeta^C\, ,~~~~ \zeta^P= \left(\begin{array}{c}e^{\frac{m}{2}r}\\0\end{array}\right)\, ,~~~~\zeta^C= e^{\frac{m}{2}r}\left(\begin{array}{c}m(t+x)e^{\frac{m}{2}r}\\e^{-\frac{m}{2}r}\end{array}\right)\,.
\eeq
where $c_1,c_2$ are constants, $\zeta^P$ is the Poincar\'{e}  (or spacetime) supercharge and $\zeta^C$ is the conformal supercharge --- together realising ${\cal N}=(1,0)$ superconformal symmetry.  Note we have taken $\gamma^{(3)}_{\mu}=(i\sigma_2,\sigma_1,\sigma_3)_{\mu}$ here. In terms of these one can define the SL(2)$_+$ doublet
\beq
\zeta^I=  \left(\begin{array}{c}\zeta^P\\ \zeta^C\end{array}\right)\, ,
\eeq
which gives rise to a  matrix of bi-linears of the form
\beq\label{eq:AdS3matribilinear}
\zeta^{I}\otimes \bar{\zeta}^{J}=\frac{1}{2}\left(\begin{array}{cc}v_1\wedge(1-u)&1+u-\frac{1}{2}v_1\wedge v_2-\text{vol}(\text{AdS}_3)\\
-1+u-\frac{1}{2}v_1\wedge v_2+\text{vol}(\text{AdS}_3)&v_2\wedge(1+u)\end{array}\right)
\eeq
where in terms of \eqref{eq:vielAdS3} the various 1-forms that appear here are
\beq
v_1= e^{2m r}(dt-dx)\, ,~~~~v_2=m^2(t+x)v_1+2m (t+x)dr+dt+dx\, ,~~~~u=dr+ m (t+x)v_1\, .
\eeq
These obey the following simple identities 
\beq
dv_1=-2m v_1\wedge u\, ,~~~dv_2=2m v_2\wedge u\, ,~~~ du=-m v_1\wedge v_2\, ,~~~v_1\wedge v_2\wedge u= 2 \text{vol}(\text{AdS}_3)\, ,
\eeq
which imply that $d(\zeta^{I}\otimes \bar{\zeta}^{J})=2m( \zeta^{I}\otimes \bar{\zeta}^{J})_2$ and $\zeta^{I}\otimes \bar{\zeta}^{J}\wedge \text{vol}(\text{AdS}_3)= -(\zeta^{I}\otimes \bar{\zeta}^{J})_3$. Further one can show that these 1-forms obey the following conditions under the Lie derivative and interior product 
\begin{align}
\nabla_{(\mu}(u_i)_{\nu)}&=0\, ,~~~ {\cal L}_{u_i} u_j= 2m f_{ijk} u_k\, ,~~~~ u_i=(\frac{1}{2}(v_1+v_2),\frac{1}{2}(v_1-v_2),u)_i\,,\nn\\[2mm]
\langle v_1,v_1 \rangle&=\langle v_2,v_2\rangle=\langle v_1,u\rangle=\langle v_2,u\rangle=0\, ,~~~~ \langle v_1,v_2\rangle=-2\, ,~~~~ \langle u,u \rangle=1\, .
\end{align}
where $f_{ijk}$ are the structure constants of SL(2), i.e.\ the Lie alebra of SL(2) is spanned by $\tau_{i}=\frac{1}{2}(i\sigma_2,\sigma_1,\sigma_3)_i$ which are such that $[\tau_{i},\tau_{j}]= f_{ijk} \tau_{k}$. We thus have that  $((v_1)^{\mu}\partial_{\mu},(v_2)^{\mu}\partial_{\mu})$ are null Killing vectors, $(u_1)^{\mu}\partial_{\mu}$ is a space-like Killing vector and $((\zeta^{I}\otimes \bar{\zeta}^{J})_1)^{\mu}\partial_{\mu}$ is a symmetric matrix  containing the three independent Killing vectors of SL(2). It also follows that
\beq
\iota_{(\zeta^{I}\otimes \bar{\zeta}^{J})_1}\text{vol}(\text{AdS}_3)=- (\zeta^{I}\otimes \bar{\zeta}^{J})_2\, .
\eeq
For the following appendix it will be useful to decompose \eqref{eq:AdS3matribilinear} as
\beq
\zeta^{I}\otimes \bar{\zeta}^{J}= \frac{1}{2}\left(\delta^{IJ}\psi^{(0)}+ (\sigma_1)^{IJ}\psi^{(1)}+i(\sigma_2)^{IJ}\psi^{(2)}+(\sigma_3)^{IJ}\psi^{(3)}\right)\, ,
\eeq
where everything here is real and $\Sigma^A=(\mathbb{I},\sigma_1,i\sigma_2 ,\sigma_3)^A$ is such that $(\Sigma^A)^{IJ}(\Sigma^B)^{IJ}=2 \delta^{AB}$. Then defining a time-like Killing vector $k^{\mu}\partial_{\mu}$ through its dual 1-form as
\beq
k= u_1= \frac{1}{2}(v_1+v_2)\, ,
\eeq
one has that
\begin{align}\label{eq:Liev}
&{\cal L}_k(\psi^{(0)}+i \psi^{(2)})=0\, ,~~~~~{\cal L}_v(\psi^{(1)}+i \psi^{(3)})=2m i (\psi^{(1)}+i \psi^{(3)})\, ,\nn\\[2mm]
&\iota_k(\psi^{(0)}+i \psi^{(2)})=i(\psi^{(0)}_2+i \psi^{(2)}_0)\, ,~~~~~ \iota_k(\psi^{(1)}+i \psi^{(3)})=i(\psi^{(1)}_1+i \psi^{(3)}_1)\, ,\nn\\[2mm]
&k\wedge(\psi^{(0)}+i \psi^{(2)})=i(\psi^{(0)}_1+i \psi^{(2)}_3)\, ,~~~~~ k\wedge(\psi^{(1)}+i \psi^{(3)})=i(\psi^{(1)}_2+i \psi^{(3)}_2)\, ,
\end{align}
i.e.\  $\psi^{0,2}$ are singlets under $v^{\mu}\partial_{\mu}$ while $\psi^{1,3}$ are charged. These expressions will be important for identifying the 7d bi-linears charged under the U(1) R-symmetry in the following appendix. Note that we also have
\beq\label{eq:map}
{\cal L}_{k}\zeta^I=m \epsilon^{IJ}\zeta^J\, ,
\eeq
providing a map between the two supercharges contained in $\zeta^I$.

\section{Detailed derivation of geometric conditions for ${\cal N}=(2,0)$ AdS$_3$ }\label{app:(2,0)}
In this appendix we will give a detailed derivation of the necessary and sufficient conditions for an AdS$_3$ solution of Type II supergravity to preserve  ${\cal N}=(2,0)$ supersymmetry; we will make use of an existing classification for totally generic Type II solutions \cite{Tomasiello:2011eb}.\\
~~\\
A solution of Type II supergravity preserving SO(2,2) in terms of an AdS$_3$ factor can in general be written in the form
\beq
ds^2= e^{2A}ds^2(\text{AdS}_3)+ds^2(\text{M}_7)\, ,~~~ H^{(10{\rm d})}= e^{3A}h_0 \text{vol}(\text{AdS}_3)+ H\, ,~~~ F= f_{\pm}+ e^{3A}\text{vol}(\text{AdS}_3)\wedge\star_7\lambda(f_{\pm})\, ,\nn
\eeq
where  $e^{2A},H,f_{\pm}$ and the dilaton $\Phi$ have support on M$_7$ alone and $e^{3A}h_0$ is a constant. Here $f_+= f_0+ f_2+f_4+f_6$ should be taken in IIA and $f_-= f_1+ f_3+f_5+f_7$ should be taken in IIB; this convention for upper/lower signs will be used throughout. For such a solution to preserve ${\cal N}=(2,0)$ supersymmetry its $d=10$ Majorana--Weyl Killing spinors should decompose as
\beq\label{eq:gen10dspinor}
\epsilon_1= \sum_{I=1}^2\zeta^I\otimes \theta_+ \otimes \chi^I_1,~~~\epsilon_2= \sum_{I=1}^2\zeta^I\otimes \theta_{\mp} \otimes \chi^I_2\, ,
\eeq
where $\zeta^I$ are a doublet of SL(2)$_+$ Killing spinors on AdS$_3$ that are Majorana, $\chi^I_{1,2}$ are independent doublets of Majorana spinors on M$_7$ and $\theta_{\pm}$ are the auxiliary vectors one always needs when decomposing an even-dimensional space in terms of two odd ones; they parameterise the $d=10$ chirality indicated by $\pm$. The astute reader will note that by identifying $\zeta^I$ appearing in \eqref{eq:gen10dspinor} with the SL(2) doublet of the previous section we are only manifestly preserving two real supercharges, where as ${\cal N}=(2,0)$ superconformal symmetry preserves four. The resolution to this naive paradox is that \eqref{eq:map} ensures that there are another two supercharges any solution consistent with \eqref{eq:gen10dspinor} must also be consistent with, i.e.\ $\epsilon_{1,2}$ as defined above, but for $\zeta^I\to \epsilon^{IJ}\zeta^J$ (that is unless $m=0$ sending AdS$_3 \to $Mink$_3$).  We shall take the $d=10$ gamma matrices to decompose as
\beq
\Gamma_{\mu}=e^{A} \gamma^{(3)}_{\mu}\otimes\sigma_3 \otimes \mathbb{I}_8\, ,~~~~~\Gamma_a= \mathbb{I}_2\otimes\sigma_1 \otimes \gamma_a\, ,
\eeq
where $\gamma^{(3)}_{\mu}$ are the real Cliff(1,2) gamma matrices of the previous appendix and $\gamma_a$ are a set of gamma matrices on M$_7$ such that $i\gamma_{1234567}=1$. The chirality matrix and intertwiner defining Majorana conjugation can then be taken to be
\beq
\hat \Gamma= -\mathbb{I}_2\otimes\sigma_2 \otimes \mathbb{I}_8\, ,~~~~ B^{(10)}=  \mathbb{I}_2\otimes \sigma_3\otimes B\, ,~~~~ B\gamma_a B^{-1}=- \gamma_a^*\, ,~~~BB^*=1\, .
\eeq
Then given that we must have $\epsilon^c_{1,2}=B^{(10)}\epsilon_{1,2}^*=\epsilon_{1,2}$ we can without loss of generality take $\zeta^I$ to be the real doublet of AdS$_3$ spinors in the previous appendix and
\beq
\theta_{\pm}=\frac{1}{\sqrt{2}}\left(\begin{array}{c}1\\\mp i\\\end{array}\right)\, .
\eeq
From this point we begin to make use of  \cite{Tomasiello:2011eb} which give necessary and sufficient geometric conditions for supersymmetry of any Type II solution, the fundamental objects are the following bi-linears in ten dimensions
\beq
K =\frac{1}{64} (\bar{\epsilon}_1 \Gamma_M \epsilon_1+ \bar{\epsilon}_2 \Gamma_M \epsilon_2)dX^M\, ,~~~\tilde{K} =\frac{1}{64} (\bar{\epsilon}_1 \Gamma_M \epsilon_1-\bar{\epsilon}_2 \Gamma_M \epsilon_2)dX^M\, ,~~~~\Psi^{(10{\rm d})}= \epsilon_1\otimes \bar{\epsilon}_2\, .
\eeq
The first necessary condition we consider is that $K^M\partial_{M}$ is a Killing vector under which both the bosonic supergravity fields and $\Psi^{(10{\rm d})}$ are singlets. Given our ansatz we find
\beq\label{eq:Killing}
K=\frac{1}{32}\big(e^{A}(\zeta^I \otimes \bar{\zeta^J})_1(\chi^{J\dag}_1\chi^I_1+\chi^{J\dag}_2\chi^I_2)-   \xi\big)\, ,~~~\tilde K=\frac{1}{32}\big(e^{A}(\zeta^I \otimes \bar{\zeta}^J)_1(\chi^{J\dag}_1\chi^I_1-\chi^{J\dag}_2\chi^I_2)-  \tilde\xi\big)\, ,
\eeq
for the real $d=7$ one-forms
\beq
\xi= -i(\chi^1_1\gamma_a\chi^2_1 \mp  \chi^1_2\gamma_a\chi^2_2)\e^a,~~~\tilde\xi= -i(\chi^1_1\gamma_a\chi^2_1 \pm  \chi^1_2\gamma_a\chi^2_2)\e^a\, ,
\eeq
where $\e^a$ is a vielbein on M$_7$. Imposing $\nabla_{(M}K_{N)}={\cal L}_K\Phi={\cal L}_KF={\cal L}_KH^{(10{\rm d})}=0$ and making use of the fact that $(\zeta^I \otimes \bar{\zeta^J})_1$is a matrix containing 1-forms dual to Killing vectors on AdS$_3$, we find the $d=7$ conditions
\begin{align}
\nabla_{(a}\xi_{b)}=0,~~~{\cal L}_{\xi}A={\cal L}_{\xi}\Phi={\cal L}_{\xi}f_{\pm}={\cal L}_{\xi}H&=0\, ,\nn\\[2mm]
 d(e^{-A}((\chi^{(I\dag}_1\chi^{J)}_1+\chi^{(I\dag}_2\chi^{J)}_2))&=0\, ,
\end{align}
necessary follow --- i.e.\ $\xi^a\partial_{a}$ is a Killing vector with respect to all bosonic supergravity fields and
\beq
\chi^{I\dag}_1\chi^{J}_1+\chi^{I\dag}_2\chi^{J}_2= e^{A}c^{IJ}_+\, ,
\eeq
for $c^{IJ}_+$ a symmetric (this follows because $\chi^I_{1,2}$ are Majorana) constant matrix such that $c^{11},c^{22}>0$. The next necessary condition we consider is $d\tilde{K}=\iota_{K}H^{(10{\rm d})}$ which gives rise to the $d=7$ conditions
\beq
d\tilde{\xi}=\iota_{\xi}H\, ,~~~~\chi^{I\dag}_1\chi^{J}_1-\chi^{I\dag}_2\chi^{J}_2= e^{-A}c^{IJ}_-\, ,~~~~~2m c^{IJ}_-=- c^{IJ}_+ e^{3A}h_0\, ,
\eeq
for $c^{IJ}_-$ another symmetric constant matrix. We can now use a constant GL(2,$\mathbb{R}$) of $\chi^{I}_{1,2}$ (which can be absorbed with a corresponding inverse transformation of $\zeta^I$ in \eqref{eq:gen10dspinor}) to fix
\beq
c^{IJ}_+= 2 \delta^{IJ}\, ,~~~~c^{IJ}_-=  \delta^{IJ} c~~~~\Rightarrow ~~~~  e^{3A}h_0=-m c\, ,
\eeq
without loss of generality. This refines \eqref{eq:Killing} as
\beq\label{eq:Killingref}
K=\frac{1}{32}\big(2e^{2A}k -   \xi\big)\, ,~~~\tilde K=\frac{1}{32}\big(c k-  \tilde\xi\big)\, ,
\eeq
for $k^{\mu}\partial_{\mu}$ the time-like Killing vector on AdS$_3$ defined in the previous appendix. Now we turn our attention to the bi-linear $\Psi^{(10{\rm d})}$ which must obey the necessary condition
\beq\label{eq:Killingconda}
(d-H^{(10{\rm d})}\wedge )(e^{-\Phi}\Psi^{(10{\rm d})})=-(\tilde{K}\wedge+\iota_{K})F\, .
\eeq
We find that the objects appearing here decompose as
\begin{align}
2\Psi^{(10{\rm d})}&= (\zeta^I\otimes \bar{\zeta}^J)_0(\Psi^{[IJ]})_{\pm}\mp(\zeta^I\otimes \bar{\zeta}^J)_3\wedge (e^{3A}\Psi^{[IJ]})_{\mp}\nn\\[2mm]
&+(\zeta^I\otimes \bar{\zeta}^J)_2\wedge (e^{2A}\Psi^{(IJ)})_{\pm}\mp (\zeta^I\otimes \bar{\zeta}^J)_1\wedge (e^{A}\Psi^{(IJ)})_{\mp}\\[2mm]
-(\tilde{K}\wedge+\iota_{K})F&=\frac{1}{32}\epsilon^{IJ}\bigg((\zeta^I\otimes \bar{\zeta}^J)_0(\tilde{\xi}\wedge+\iota_{\xi})f+ e^{3A}(\zeta^I\otimes \bar{\zeta}^J)_3\wedge (\tilde{\xi}\wedge+\iota_{\xi})\star_7  \lambda f_{\pm}\bigg)\nn\\[2mm]
&+ \frac{1}{16}e^{3A}\delta^{IJ}\bigg[-\frac{c}{2}(\zeta^I \otimes \bar{\zeta}^J)_1\wedge f_{\pm} +(\zeta^I \otimes \bar{\zeta}^J)_2\wedge \star_7 \lambda f_{\pm}\bigg]\nn\, ,
\end{align}
where we define the $d=7$ matrix bi-linear
\beq
\Psi^{IJ} \equiv \chi_1^I\otimes \chi_2^{J\dag}\, .
\eeq
Plugging this into \eqref{eq:Killingconda} yields the $d=7$ differential bi-linear constraints
\begin{subequations}
\begin{align}
&d_{H}(e^{A-\Phi}\Psi^{(IJ)}_{\mp})=\mp \frac{c}{16}\delta^{IJ}f_{\pm}\, ,\label{eq:apbps1}\\[2mm]
&d_{H}(e^{2A-\Phi}\Psi^{(IJ)}_{\pm})\mp 2m e^{A-\Phi}\Psi^{(IJ)}_{\mp}= \frac{1}{8}e^{3A}\star_7\lambda f_{\pm} \delta^{IJ}\, ,\label{eq:apbps2}\\[2mm]
&d_{H}(e^{-\Phi}\Psi^{[IJ]}_{\pm})=\frac{1}{16}\epsilon^{IJ}\big(\tilde{\xi}\wedge+\iota_{\xi}\big) f_{\pm}\, ,\label{eq:apbps3}\\[2mm]
&d_{H}(e^{3A-\Phi}\Psi^{[IJ]}_{\mp})=\mp e^{3A-\Phi}h_0\Psi^{[IJ]}_{\pm}\pm \frac{1}{16}\epsilon^{IJ}\big(\tilde{\xi}\wedge +\iota_{\xi}\big)e^{3A}\star_7 \lambda f_{\pm}\, ,\label{eq:apbps4}
\end{align}
\end{subequations}
where we note that the (11) and (22) components of \eqref{eq:apbps1}-\eqref{eq:apbps2} reproduce the differential ${\cal N}=(1,0)$ conditions presented in \cite{Macpherson:2021lbr}, as they should. The conditions derived thus far are not sufficient for supersymmetry to hold, for that one must also solve the pairing constraints, namely  (3.1c)-(3.1d) of \cite{Tomasiello:2011eb}. Generically these are the hardest conditions to deal with, however in this case we can rely on earlier AdS$_3$ work for ${\cal N}=(1,0)$ solutions \cite{Dibitetto:2018ftj},\cite{Passias:2019rga},\cite{Macpherson:2021lbr} (respectively the original work, first to use these conventions and first to generalise to $c\neq 0$) which informs us that these conditions are implied by 
\beq
(\Psi^{11}_{\mp},f_{\mp})_7=(\Psi^{22}_{\mp},f_{\mp})_7=\mp\frac{1}{2}\left(m+\frac{1}{4}e^{-A}ch_0\right)e^{-\Phi}\text{vol}(\text{M}_7)\, ,
\eeq
however one can show that the difference of these conditions is actually implied by the trace of \eqref{eq:apbps2} --- a similar outcome was found for ${\cal N}=(1,1)$ AdS$_3$ and the steps to show this are analogous (see appendix C of \cite{Macpherson:2021lbr}). Further one can show that $(\Psi^{(1)}_{\mp},f_{\mp})_7=(\Psi^{(3)}_{\mp},f_{\mp})_7=0$ allowing one to write the pairing constraints in a covariant fashion as 
\beq
(\Psi^{(IJ)}_{\mp},f_{\mp})_7=\mp\frac{1}{2}\delta^{IJ}\left(m+\frac{1}{4}e^{-A}ch_0\right)e^{-\Phi}\text{vol}(\text{M}_7)\, ,
\eeq
where only the trace of this contains non-trivial information. We have now derived a necessary and sufficient set of conditions for ${\cal N}=(2,0)$ supersymmetry, these can however be refined somewhat: It is well known that the ${\cal N}=(2,0)$ AdS$_3$ solutions come equipped with a U(1) R-symmetry under which the spinors $\chi^I_{1,2}$ should be charged --- having established that $\xi^a\partial_{a}$ is necessarily a Killing vector, clearly it is this that should be identified with that U(1). Indeed a consequence of supersymmetry is that ${\cal L}_K\epsilon_{1,2}=0$, and since $K$ is spanned by $k$ and $\xi$ with $\zeta^I$ transforming non triviality under the former, clearly $\chi^I_{1,2}$ must transform under $\xi^a\partial_{a}$ for \eqref{eq:gen10dspinor} to be consistent. We find
\beq
{\cal L}_{\xi}\chi_{1,2}^I= -2m \epsilon^{IJ}\chi^J_{1,2}\, ,
\eeq
i.e.\ $\chi^I_{1,2}$ are SO(2) doublets as expected. As the matrix bi-linear $\Psi^{IJ}$ is a tensor product of SO(2) doublets it should decompose into irreducible representations of SO(2) as $\textbf{2}\otimes\textbf{2}=\textbf{1}\oplus \textbf{1}\oplus \textbf{2}$, as such $\Psi^{IJ}$ should contain both singlet and doublet contributions. To see this it is helpful to decompose
\beq
\Psi^{IJ}= \frac{e^{A}}{2}\left(\delta^{IJ}\Psi^{(0)}+\sigma_1^{IJ}\Psi^{(1)}+i \sigma_2^{IJ}\Psi^{(2)}+\sigma^{IJ}_3\Psi^{(3)}\right)\, ,~~~ \Psi^{(0,1,2,3)}=\Psi^{(0,1,2,3)}_++i \Psi^{(0,1,2,3)}_-\, ,
\eeq
where $\Psi^{(0,1,2,3)}_{\pm}$ are real. We mentioned before that $\Psi^{(10{\rm d})}$ should be a singlet with respect to $K$,  given that the  AdS$_3$ bi-linears transform non-trivially under $k$ as in \eqref{eq:Liev}, it follows that
\beq\label{eq:liecond}
{\cal L}_K\Psi^{(10)}=0~~~~\Rightarrow~~~~{\cal L}_{\xi}\Psi^{(0)}={\cal L}_{\xi} \Psi^{(2)}=0\, ,~~~{\cal L}_{\xi}(\Psi^{(1)}+i \Psi^{(3)})=- 4 i m  (\Psi^{(1)}+i \Psi^{(3)})\, ,
\eeq
so it is only $\Psi^{(1,3)}$ that are charged under the U(1) R-symmetry. Another useful condition that follows when supersymmetry holds is that $(\iota_{K}+\tilde{K}\wedge)\Psi^{(10{\rm d})}=0$, which one can show implies the following conditions on the $d=7$ bi-linears 
\begin{align}
(\iota_{\xi}+\tilde{\xi}\wedge)(\Psi^{(0)}_{\mp}+i \Psi^{(2)}_{\mp})&=\mp i ce^{-A}(\Psi^{(0)}_\pm+i \Psi^{(2)}_{\pm})\, ,\nn\\[2mm]
(\iota_{\xi}+\tilde{\xi}\wedge)(\Psi^{(1)}_{\mp}+i \Psi^{(3)}_{\mp})&=\mp 2 i e^{A}(\Psi^{(1)}_\pm+i \Psi^{(3)}_{\pm})\, ,\nn\\[2mm]
(\iota_{\xi}+\tilde{\xi}\wedge)(\Psi^{(0)}_{\pm}+i \Psi^{(2)}_{\pm})&=\pm 2i e^{A}(\Psi^{(0)}_{\mp}+i \Psi^{(2)}_{\mp})\, ,\nn\\[2mm]
(\iota_{\xi}+\tilde{\xi}\wedge)(\Psi^{(1)}_{\pm}+i \Psi^{(3)}_{\pm})&=\pm i ce^{-A}(\Psi^{(1)}_{\mp}+i \Psi^{(3)}_{\mp})\, .
\end{align}
Additionally $\langle K,\tilde{K}\rangle=0$ must hold, as $K, \tilde{K}$ are proportional to the sum and difference of 2 null vectors, this implies
\beq
\langle \xi,\tilde{\xi}\rangle= 2 c\, .
\eeq
From these identities one can show that \eqref{eq:apbps4} and the $\Psi^{(0)}_{\pm}$ dependent contribution to \eqref{eq:apbps1} are implied by the other conditions in general.\\
~\\
This concludes our derivation of the necessary and sufficient geometric conditions for ${\cal N}=(2,0)$ AdS$_3$ solutions of Type II supergravity, we summarise our results and present them in a compact fashion in section \ref{sec:SUSY}.

\section{SU(2)-structure and torsion classes in six dimensions}\label{app:torsions}
An SU(2)-structure in six dimensions is defined by a $(1,0)$-form $z$ a real $(1,1)$-form $j_2$ and a $(2,0)$-form $\omega_2$ which obey the relations
\beq\label{eq:SU(2)conditions}
j_2\wedge \omega_2=0\, ,~~~ j_2\wedge j_2=\frac{1}{2}\omega_2\wedge \overline{\omega}_2\, ,~~~\omega_2\wedge \omega_2=0\, ,
\eeq 
with $z$ orthogonal to $j_2$ and $\omega_2$. As such an SU(2)-structure indicates an obvious decomposition of the $d=6$ metric as
\beq
ds^2(\text{M}_6)=  \frac{i}{2}z \overline{z}+ ds^2(\text{M}_4)\, .
\eeq
There always exists a canonical frame  $(\e^1,\e^2,\e^3,\e^4)$ of M$_4$  such that
\beq
j_2= \e^1\wedge \e^2+ \e^3\wedge \e^4,~~~~\omega_2=(\e^1+i \e^2)\wedge (\e^3+i \e^4)~~~\Rightarrow~~~\frac{1}{2}j_2\wedge j_2=\e^{1234}=\text{vol}(\text{M}_4)\, ,
\eeq 
which makes clear that $(j_2,\text{Re}\omega_2,\text{Im}\omega_2)$ span the real self-dual 2-forms on M$_4$.\\
~\\
In the main text it will be useful to know the torsion classes for an SU(2)-structure in six dimensions, these can be computed with group theory given that the torsion classes should form irreducible representations of SU(2), here we will take a different approach and exploit a canonical frame in $d=6$.

The aim is to decompose the exterior derivatives of $(z,j_2,\omega_2)$ in terms of objects with useful properties under the wedge product and hodge dual -- these will in fact turn out to be irreducible representations of SU(2), but this fact is somewhat auxiliary to this usefulness. The first step is to introduce some complex primitive (i.e.\ $j_2\wedge T_i=0$) (1,1)-forms $T_i$, holomorphic 1-forms $V_i$ and complex functions $S_i$ (here $i \in \mathbb{N}$). In the canonical frame
these are expressed in a basis of the following
\beq
V_i:~ (\e^1+i \e^2, \e^3+i \e^4),~~~~~ T_i:~ (\e^1\wedge \e^2-\e^3\wedge \e^4,\e^1\wedge \e^3+\e^2\wedge \e^4,\e^1\wedge \e^4-\e^2\wedge \e^3)\, .
\eeq
It is then a simple matter to confirm that one can express generic $n$-forms on M$_4$ $X^{(4)}_n$ in a basis of the forms introduced so far as
\beq
X^{(4)}_1= V_1+\overline{V}_2\, ,~~~X^{(4)}_2= S_1\omega_2+S_2\overline{\omega}_2+S_3 j_2+T_1\, ,~~~~X^{(4)}_3= (V_3+\overline{V}_4)\wedge j_2\, ,~~~~X^{(4)}_4= S_5 j_2\wedge j_2\, .
\eeq 
Relevant to the torsion class are generic complex 2 and 3-forms on M$_6$, which can thus be decomposed as
\begin{align}
X^{(6)}_2&= S_1\omega_2+S_2\overline{\omega}_2+S_3 j_2+T_1+ S_4 z\wedge \overline{z}+ (V_1+\overline{V}_2)\wedge z+ (V_3+\overline{V}_4)\wedge \overline{z}\, ,\\[2mm]
X^{(6)}_3&=(V_5+\overline{V}_6)\wedge j_2 +z\wedge(S_5\omega_2+S_6\overline{\omega}_2+S_7 j_2+T_2) +\overline{z}\wedge (S_8\omega_2+S_9\overline{\omega}_2+S_{10} j_2+T_3)\nonumber\\
&+ (V_7+\overline{V}_8)\wedge z\wedge \overline{z}\nn\, .
\end{align}
However, $dj_2,d\omega_2$ are not generic, they must also be consistent with the exterior derivatives of \eqref{eq:SU(2)conditions} which forces some terms to be equal. We find in general that the torsion classes take the form
\begin{align}
dz&= S_1\omega_2+S_2\overline{\omega}_2+S_3 j_2+T_1+ S_4 z\wedge \overline{z}+  z\wedge W_1+  \overline{z}\wedge W_2\, ,\nn\\[2mm]
d\omega_2&=(V_1+\overline{V}_2)\wedge j_2 +z\wedge(S_5\omega_2+S_{6} j_2+T_2) +\overline{z}\wedge (S_7\omega_2+S_{8} j_2+T_3)+ i (\iota_{\overline{V}_3}\omega_2)\wedge z\wedge \overline{z}\, ,\nn\\[2mm]
dj_2&=V_4\wedge j_2+V_3\wedge z\wedge \overline{z}+z\wedge(-\frac{1}{2}\overline{S}_{8}\omega_2-\frac{1}{2}S_{6}\overline{\omega}_2+\frac{1}{2}(S_5+\overline{S}_7)j_2+T_4)+ \text{c.c}\, ,
\end{align}
where $W_i$ are simply complex 1-forms on M$_4$ that can each be expressed in terms of two holomorphic 1-forms if we wish. Note this gives the correct counting 8 scalars, 8 holomorphic 1-forms and 4 primitive (1,1)-forms.\\
~\\
As promised, by making use of the canonical frame one can compute many nice identities the SU(2)-structure forms and their torsion classes must obey --- for instance
\begin{align}
& T_i\wedge j_2= T_i \wedge \omega_2= V_i\wedge \omega_2=0,\nn\\[2mm]
&\star_6V_i=\frac{1}{2}V_i\wedge j_2\wedge z\wedge \overline{z},~~~~\star_6(V_i\wedge j_2)=\frac{1}{2}V_i\wedge z\wedge \overline{z},\nn\\[2mm]
& \star_6 j_2=\frac{i}{2} j_2\wedge z\wedge \overline{z},~~~~\star_6 \omega_2=\frac{i}{2}\omega_2\wedge z\wedge \overline{z},~~~~\star_6 T_i=-\frac{i}{2}T_i\wedge z\wedge \overline{z},\nn\\[2mm]
&(\iota_{\overline{V}_i}\omega_2)\wedge j_2=-i \overline{V}_i\wedge \omega_2\, ,~~~~(\iota_{\overline{V}_i}\omega_2)\wedge \overline{\omega}_2=4i \overline{V}_i\wedge j_2\, ,
\end{align}
which we make use of in the main text.

%
%    \bibliographystyle{JHEP}
%    
%    
%    
%    \bibliography{nullbib}
%    
%    
%    \end{document}

%%%%%%%%%%%%%%%%%%%%%%%%%%%%%%%%%%%%%%%%%

\end{document}